\documentclass[sigconf,nonacm]{acmart}

\usepackage[T1]{fontenc}
\usepackage[english]{babel}
\usepackage{booktabs}
\usepackage{tabularx}
\usepackage{booktabs}
\usepackage{multirow}
\usepackage{graphicx}
\usepackage{comment}

\begin{document}

\title{Day in the Life of RIPE Atlas: Operational Insights and Applications in Network Measurements}

\author{Yevheniya Nosyk}
\affiliation{
  \institution{KOR Labs}
  \country{France}
}

\author{Malte Tashiro}
\affiliation{
  \institution{IIJ Research Laboratory / SOKENDAI}
  \country{Japan}
}

\author{Qasim Lone}
\affiliation{
  \institution{RIPE NCC}
  \country{The Netherlands}
}

\author{Robert Kisteleki}
\affiliation{
  \institution{RIPE NCC}
  \country{The Netherlands}
}

\author{Andrzej Duda}
\affiliation{
  \institution{Université Grenoble Alpes}
  \country{France}
}

\author{Maciej Korczyński}
\affiliation{
  \institution{Université Grenoble Alpes}
  \country{France}
}

\begin{abstract}
Network measurement platforms are increasingly popular among researchers and operators alike due to their distributed nature, simplifying measuring the remote parts of the Internet. RIPE Atlas boasts over 12.9~K vantage points in 178 countries worldwide and serves as a vital tool for analyzing anycast deployment, network latency, and topology, to name a few. Despite generating over a terabyte of measurement results per day, there is limited understanding of the underlying processes. This paper delves into one day in the life of RIPE Atlas, encompassing 50.9~K unique measurements and over 1.3 billion results. While most daily measurements are user-defined, it is built-ins and anchor meshes that account for 89\% of produced results. We extensively examine how different probes and measurements contribute to the daily operations of RIPE Atlas and consider any bias they may introduce. Furthermore, we demonstrate how existing measurements can be leveraged to investigate censorship, traceroute symmetry, and the usage of reserved address blocks, among others. Finally, we curate a set of recommendations for researchers using the RIPE Atlas platform to foster transparency, reproducibility, and ethics.
\end{abstract}

\maketitle

\section{Introduction\label{sec-introduction}}

The Internet is the biggest network of networks in the world, interconnecting more than 82.3~K IPv4/IPv6 autonomous systems that advertise over 1 million BGP routing prefixes~\cite{routeviews}. Given this unprecedented scale, it is no surprise that the Internet is constantly affected by network outages that stem from cyberattacks, natural disasters, hardware or software failures, as well as human errors. Such events may have a global impact and far-reaching consequences for millions of end users. 

One of the prominent network outages happened in 2016 when the Mirai botnet attacked the Dyn Domain Name System (DNS) operator. It resulted in rendering unavailable some of the most popular websites, such as those of Amazon, GitHub, and Netflix, to name a few~\cite{mirai}. More recently, Facebook and its associated services, Instagram and WhatsApp, went offline when one of the ``routine maintenance jobs'' disconnected the Meta network hosting DNS servers from the Internet~\cite{fb-outage}. It is estimated that the outage cost the company about \$65 million~\cite{forbes-estimation}. The understanding and mitigation of such disruptions is more efficient when one can accurately measure how far the problem propagated and which parts of the Internet were affected. It is also crucial to proactively identify any vulnerability or misconfiguration that may become an issue in the future. 

Network measurements can be conducted from external or internal vantage points, each offering unique insights into network performance. External vantage points evaluate network reachability and end-to-end performance, identifying issues such as connectivity problems and high latency. However, they often lack the depth needed to uncover the root causes of these issues. In contrast, internal vantage points are essential for gaining a comprehensive understanding of network operations. They enable detailed monitoring of traffic patterns and facilitate the identification of internal issues, such as routing misconfigurations and bandwidth bottlenecks.

To fully leverage the strengths of internal vantage points, developing a distributed infrastructure that supports a variety of custom measurements across multiple networks is crucial. Such a system would not only streamline automation and simplify result retrieval but also significantly enhance overall network visibility. Examples of such systems are proxy networks, advertising tens of millions of vantage points available to their clients. Researchers previously used them to test the capabilities of DNS resolvers~\cite{longitudinal}, DNS time-to-live (TTL) violations~\cite{ttl-violations}, and traffic manipulations~\cite{tunneling}. Yet, valid concerns have been raised as to whether participating IP addresses are sourced ethically. In particular, Mi~et~al.~\cite{resident-evil} found some of the residential proxies to be compromised Internet of Things devices. NLNOG RING~\cite{ring} is another initiative introduced by the Netherlands Network Operator Group. Any operator joining the measurement network with its own virtual machine gains shell access to all the participating servers. However, the platform is only open to participants.

RIPE Atlas stands out as one of the most widely known Internet measurement platforms, appreciated by researchers and operators alike. Powered by a network of volunteers running small measurement devices (probes), it constantly collects data about the topology, connectivity, and reachability of the global Internet. The platform is open for anyone to launch custom measurements, provided a user possesses a sufficient amount of RIPE Atlas credits. If in need, researchers can request them by contacting the RIPE Atlas team directly. For that reason, the platform has been extensively used in the research community to study network latency~\cite{remote-peering,reassessing,latency-characteristics,last-mile,divided-edge,painter,middle-east,private-wan,inter-city}, topology~\cite{flat-internet,igdb,reverse-traceroute,regional-topology,hoplets,youtube}, and anycast~\cite{old-but-gold,manycast2,taming,regional-anycast,anycast-context,anycast-agility,cdn-routing,bidirectional-anycast} among many other topics.
 
By 2024, RIPE Atlas has grown into a large-scale network of 12.9~K probes, spread over 178 countries and 4~K ASes. Since its initial launch in 2010, the platform has generated almost one petabyte of data~\cite{reducing-footprint}, with a rate of 1.3 billion measurement results per day in early 2024. Such a scale made the Atlas team reflect on how to best store all the invaluable data~\cite{data-retention}. However, there has not been a study that systematically analyzes the measurements being run and the phenomena they can reveal. Our paper fills this research gap and provides an extensive overview of the RIPE Atlas platform, with its building blocks, measurements, and use cases. Overall, our contributions are as follows:

\begin{itemize}

    \item We analyze one day in the life of RIPE Atlas, comprising 50.9~K measurements and 1.3 billion results generated by 12.9~K probes. We uncover that anchors produce almost 70\% of daily results despite regular users launching the highest number of individual measurements. 

    \item We present several exploratory case studies that leverage RIPE Atlas data to provide insights into network behavior and infrastructure. By analyzing a single day's worth of RIPE Atlas data, we highlight the unique value of this daily snapshot dataset in uncovering patterns and trends that inform network performance and reliability. Our findings demonstrate the practical applications of these data in real-world scenarios, offering valuable implications for network operators and researchers alike.

    \item Based on the key findings of this paper and the discussions with the RIPE Atlas Team, we curate the list of best practices for all researchers wishing to use the platform. We call for transparency, reproducibility, ethical conduct, and encourage thorough exploration of RIPE Atlas data to uncover valuable insights for the network community.

\end{itemize}

The rest of the paper is organized as follows. Section~\ref{sec-background} describes the inner workings of the RIPE Atlas, while Section~\ref{sec-related-work} illustrates the use of the platform by researchers. Section~\ref{sec-dataset} presents the day-in-the-life dataset that we analyze in the remainder of the paper. Section~\ref{sec:day-in-the-life} discusses anchoring, built-in, and user-defined measurements. Section~\ref{sec-call} proposes a set of recommendations for a better use of RIPE Atlas, Section~\ref{sec-ethics} deals with ethics, and Section~\ref{sec-conclusions} concludes the paper.

\section{Background on RIPE Atlas\label{sec-background}}

This section provides an overview of the RIPE Atlas measurement platform. We first describe the overall system design and its main building blocks. We then discuss the types of supported measurements and how one can launch them.

\subsection{Building Blocks}

The core element of the RIPE Atlas platform is a \textit{probe}, capable of running in its host network and executing various types of Internet measurements. Probes come in two flavors---hardware and software. The former are small low-power devices (Lantronix XPort Pro, TL-MR 3020, or NanoPi NEO Plus2, depending on the version~\cite{iot-atlas}) following the install-and-forget principle, thus requiring little to no maintenance from the probe owner. On the contrary, software probes are set up and updated independently on the existing infrastructure (e.g., virtual machines or physical servers). RIPE NCC receives applications from potential probe hosts and accepts those that would contribute to the better coverage of the platform. In return, probe owners gain \textit{credits} that can be used to run custom measurements. As of February 2024, there are 12.9~K connected probes.

\textit{Anchor} is an enhanced probe, providing two main services: i) executing a large number of measurements and ii) being measurement targets for other probes and anchors. Accomplishing these two goals raises higher requirements for the host network and the underlying hardware. Anchors are expected to be highly available, so they are not suitable for small home or office networks behind firewalls. Instead, Internet Service Providers (ISPs), Internet eXchange Points (IXPs), or big cloud providers are natural candidates for hosting such systems. Note that, unlike regular probes, anchors are explicitly required to respond to pings and traceroutes. To reward the owners for operating anchors, they earn ten times more credits than regular probe hosts.

Originally, all the anchors were meant to be hardware devices, and hosts were committed to purchasing the specific hardware approved by the Atlas team. Non-profit organizations could additionally benefit from several rounds of sponsorship, helping to set up anchors in underrepresented networks~\cite{2018-campaign}. However, RIPE NCC relaxed the hardware requirement in 2018 and introduced Virtual Machine Anchors that contributed to a large expansion of the infrastructure~\cite{atlas-400-anchors}. As of February 2024, 810 anchors are connected to the RIPE Atlas platform. They constitute a full mesh and perform a series of measurements between themselves to serve as a baseline for the state of Internet connectivity. 

\subsection{Measurements}

The RIPE Atlas \textit{measurement} is the key unit of work performed by all the probes and anchors. Every device comes with a series of \textit{built-in} measurements performed towards well-known targets, predefined by the Atlas team. At the same time, one can launch custom \textit{user-defined} measurements via the web interface\footnote{https://atlas.ripe.net} or the API.\footnote{\url{https://atlas.ripe.net/docs/apis/rest-api-manual/}} Each measurement is defined by its type, participating probes/anchors, targets, and timing.

Built-in measurements~\cite{builtin} generate an important body of data about global Internet connectivity. They are executed on all the connected probes every 4 minutes to once per day, some of them up and running since 2010. Every 240 seconds, pings are sent to all 13 DNS root servers and 7 measurement targets from the RIPE Atlas infrastructure. Traceroutes are executed towards the same destinations every 30 minutes in addition to topology scans running every 15 minutes. DNS measurements are duplicated over both TCP and UDP transport protocols, with each target receiving \texttt{SOA} DNS and 4 \texttt{CHAOS}-class \texttt{TXT} requests for \texttt{version.bind}, \texttt{hostname.bind}, \texttt{id.server}, \texttt{version.server}. In addition, probes resolve some of the popular and random domain names. The frequency of DNS measurements greatly varies depending on the particular query sent. 

The Transport Layer Security (TLS) certificates of \url{www.ripe.net} and \url{atlas.ripe.net} are retrieved once per day, and the HTTP measurements towards \url{http://www.ripe.net/favicon.ico} and \url{http://ip-echo.ripe.net/} are executed every 24 and 1 hour, respectively. If supported by the host network, all the measurements are run in IPv4 and IPv6 address spaces. Moreover, all the probe connection and disconnection events are also logged to give an overview of the probe uptime. Built-in measurements are public by default, and the results can be freely accessed by any interested party. 

RIPE Atlas anchors form a distinct group of targets for so-called anchoring measurements. Every anchor receives a ping, a traceroute, and an HTTP GET request from all the other anchors (thus forming a full mesh) and a subset of regular RIPE Atlas probes. The measurements are run every 4 to 30 minutes to create a reliable picture of the global Internet connectivity. Whenever available, all the measurements are run over IPv4 and IPv6. Note that the list of anchoring measurements is not static and gets updated every time a new anchor is added to the mesh.

One may need to perform a more fine-grained measurement, e.g., resolve a custom domain name or ping a particular destination. In this case, built-ins would be of little help. Instead, any registered user can launch a custom measurement, provided they have a sufficient amount of credits. There are currently 6 supported measurement types, including ping, traceroute, DNS, NTP, TLS, and HTTP~\cite{user-defined}. Note that the last one is restricted to target only RIPE Atlas anchors. By default, the web interface proposes to run a measurement on 50 random probes located worldwide. It is also possible to restrict the participants to a specific AS, an IP prefix, a country, or a region. 

If known, a particular probe ID can also be specified as well as an existing measurement ID to reuse the same set of participants. Each measurement can either be one-off or recurring (e.g., every 5 minutes). Based on its definition, each measurement has an associated cost in credits, proportional to the load placed on the probes themselves and the RIPE Atlas infrastructure in general. Finally, certain limits in place prevent a single user from abusing the platform. One cannot run more than 100 measurements simultaneously, use more than 1000 probes per measurement, and spend more than 1~M credit points per day. All the special requests to bypass the quotas and limitations are considered by the Atlas team on a case-by-case basis.

\section{Related Work\label{sec-related-work}}

Since the launch of RIPE Atlas in 2010 and its introduction to the research community in 2015~\cite{ripe-paper}, the platform has been widely used in more than a thousand scientific publications~\cite{scholar}. To tackle this amount of related work in the field efficiently, we adopt the approach of Scheitle~et~al.~\cite{internet-top-lists}. Specifically, we have surveyed all the papers published in the past 5 years (2019-2023) at relevant conferences in the areas of Internet measurements (PAM, ACM IMC, TMA), networking (ACM CoNEXT, ACM/IRTF ANRW, IEEE INFOCOM, ACM SIGCOMM), and security (USENIX Security, IEEE S\&P, IEEE Euro S\&P, ACM CCS, NDSS). We looked for publications that relied on RIPE Atlas in their methodology and selected 79 papers for further analysis. 

Given that the RIPE Atlas platform is mostly considered to be a network monitoring tool, it is of no surprise that 89\% of selected publications appeared in Internet measurement and networking venues. Consequently, it was largely underrepresented in the security community. Researchers mostly performed traceroute, DNS, and ping experiments, only occasionally referring to other scan types. RIPE Atlas was used to study anycast deployment~\cite{old-but-gold,manycast2,taming,regional-anycast,anycast-context,anycast-agility,cdn-routing,bidirectional-anycast}, latency~\cite{remote-peering,reassessing,latency-characteristics,last-mile,divided-edge,painter,middle-east,private-wan,inter-city}, topology~\cite{flat-internet,igdb,reverse-traceroute,regional-topology,hoplets,youtube}, and IP geolocation~\cite{replication-geolocation,reverse-geolocation,ipvseeyou,on-the-accuracy,iot-location}. Apart from those, researchers tackled a broad number of networking topics, including anomalies and outages~\cite{passive-analysis,colocation-disruptions,perceiving-anomalies}, Resource Public Key Infrastructure (RPKI)~\cite{disco,rpki-validation,rovista,time-of-flight,rov-mi,revisiting-rpki}, router fingerprinting~\cite{snmp,illuminating}, Border Gateway Protocol (BGP)~\cite{curtain,zombies,multipath}, traceroute analysis~\cite{metatrace,rrr,timeseries}, IPv6~\cite{6hit,debogonising}, routing loops~\cite{loops}, submarine connectivity~\cite{unintended,out-of-mind}, and satellite~\cite{starlink,dissecting}. They used both built-in and user-defined measurements with a slight preference towards the latter.

DNS measurements constituted another important part of related work covering a wide range of topics such as DNS Security Extensions (DNSSEC)~\cite{roll,agility}, QNAME minimization~\cite{first-look,second-look}, caching~\cite{cache-outside,trufflehunter,cache-me,negative-caching}, DNS-over-TCP~\cite{dns-over-tcp}, encryption~\cite{dot-edge,doh-world}, EDNS(0) Client Subnet~\cite{apple-relay,ecs-behavior,edns-adoption}, fragmentation~\cite{falling-bits}, cyclic dependencies~\cite{tsuname}, lame delegations~\cite{disagree}, and infrastructure~\cite{peek,managed-dns}. However, the need to query custom domain names implied setting up user-defined measurements. 

The remaining measurement types are a small fraction of related work. Notably, different RIPE Atlas measurement types (TLS, DNS, traceroute) allowed researchers to analyze censorship from multiple angles. For example, they examined DNS interception~\cite{intercept-and-inject,home-hijacking,cryptocurrency}, rogue root certificates~\cite{kazakhstan}, censorship circumvention~\cite{netshuffle}, and more broadly, the exposure to traffic observation and tampering~\cite{selective-tampetring}. The latter two relied on TLS measurements that were rarely seen in other existing work. Finally, probe (dis)connection events were used~\cite{blocklisting,count-bots} to identify IP address reuse.  

Given the number of measurements running on the platform and the associated load on RIPE Atlas probes, valid concerns were raised about the accuracy of the results. Holterbach~et~al.~\cite{interference} found that i) measurements running concurrently on the same probe may increase the delay by milliseconds and ii) measurements on different probes can become desynchronized despite being launched at the same time. Bajpai~\cite{lessons-learned} further summarized the potential issues that may be encountered by RIPE Atlas users, such as rate limits, per-hop latency aggregation, and a biased probe distribution across ASes/countries. Researchers specifically highlighted that vantage point locations may significantly influence the obtained measurement results~\cite{bias-in-platforms,metis}. 

Our analysis of the related work shows several patterns of RIPE Atlas usage among researchers. First, measurements are skewed towards the main three types (namely traceroutes, DNS, and ping) while others remain underrepresented. Second, user-defined measurements are used in more than half of the cases because built-ins cannot address all the research questions. Finally, security researchers rely considerably less on RIPE Atlas than the networking community.

\section{Day in the Life of RIPE Atlas\label{sec-dataset}}

This section provides an overview of one day in the life of RIPE Atlas. We carefully select a 24-hour window to analyze the active probes, anchors, and measurements in which they participated.

\begin{figure}[t]
    \centering
    \includegraphics[width=0.75\columnwidth]{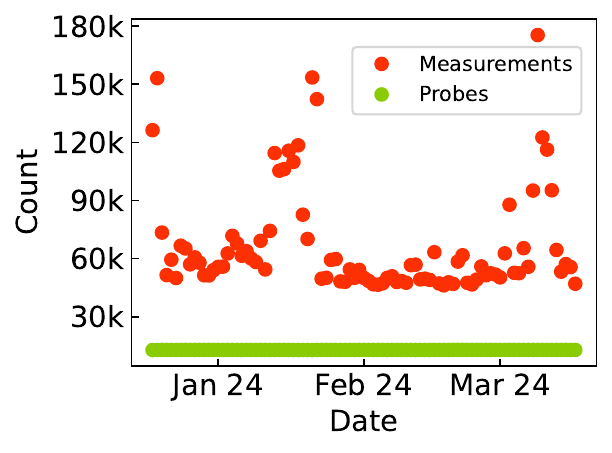}
    \caption{Ongoing measurements and active RIPE Atlas probes per day between Jan 1, 2024 and Mar 31, 2024.}
    \Description{Ongoing measurements and active RIPE Atlas probes per day between Jan 1, 2024 and Mar 31, 2024.}
    \label{fig-dataset-3m}
\end{figure}

\subsection{Dataset}

When describing one day in the life of RIPE Atlas, it is crucial to evaluate whether the chosen dataset is a representative snapshot of the system operations. We assess it using two distinct metrics: the number of ongoing measurements and the number of connected probes per day. 

Considering the major growth of the platform since 2010, we refrain from the naive approach of computing the mean values over the past 14 years. Instead, we focus on a more recent period of time spanning the first three months of 2024. Figure~\ref{fig-dataset-3m} presents the number of connected probes and ongoing measurements between January and March 2024. Overall, we see a stable distribution of probes, fluctuating between 12.8~K and 12.9~K per day with a mean of 12,888. However, trends differ for ongoing measurements, where the numbers range from 46~K to 175~K. The mean of such a distribution will inevitably be skewed towards a few outliers. Therefore, we compute its median of 56~K measurements per day. Finally, we choose the date for which the numbers of ongoing measurements and connected probes fall within the range of the median absolute deviation, set to 7~K for measurements and 14 for probes. One of the days that satisfies these criteria is February 21, 2024. We note that while the analysis of data for any short period is not necessarily illustrative of the overall system use, we argue that the chosen day is a representative snapshot of the platform operation in 2024. We download all the active measurements during this period, all the generated results, and the probe metadata using the RIPE Atlas API.

\begin{figure}[t]
    \centering
    \begin{minipage}[b]{0.45\columnwidth}
        \includegraphics[width=\textwidth]{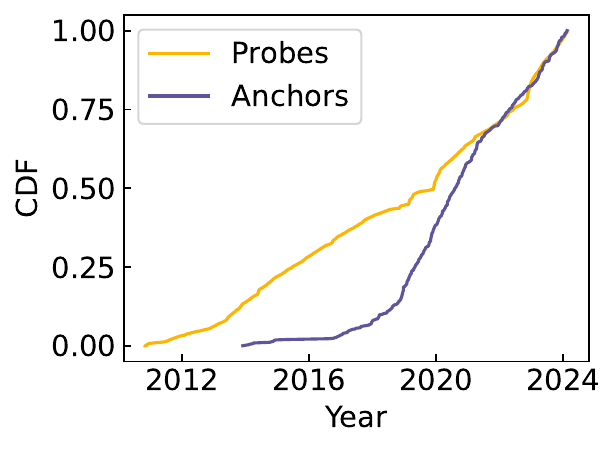}
        \caption{Cumulative distribution of dates of the first connection of probes and anchors.}
        \Description{Cumulative distribution of dates of the first connection of probes and anchors.}
        \label{fig:probe-first-connected}
    \end{minipage}
    \hfill
    \begin{minipage}[b]{0.45\columnwidth}
        \includegraphics[width=\textwidth]{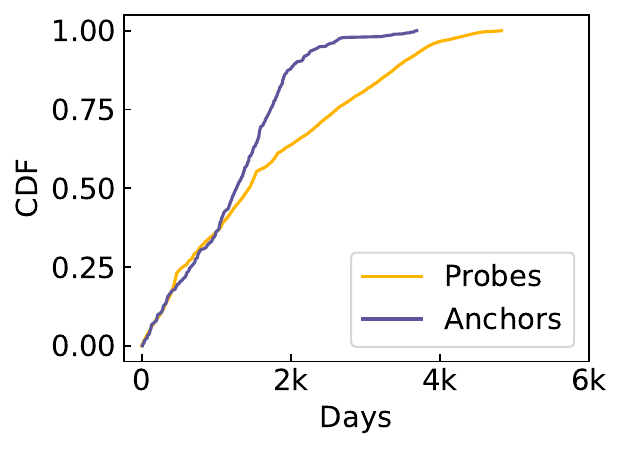}
        \caption{Cumulative distribution of total uptimes in days for probes and anchors.}
        \Description{Cumulative distribution of total uptimes in days for probes and anchors.}
        \label{fig:probe-total-uptime}
    \end{minipage}
\end{figure}

\subsection{Probes and Anchors}

On the analyzed day, 36.9~K measurement devices (probes and anchors) are known to the RIPE Atlas platform. More than 20~K of them are abandoned---they have not been connected to the system for at least several months. Almost 4~K probes/anchors are temporarily disconnected, for example, possibly due to transient network failures, probe hardware not functioning properly~\cite{troubleshooting}, or owners unplugging the device for any reason. The remaining 12,082 probes and 810 anchors are active, forming the basis for our subsequent analysis.

\subsubsection{First Connection Dates and Uptimes.} Figure~\ref{fig:probe-first-connected} shows the cumulative distribution of the dates of the first connection of the currently active probes and anchors. In general, probes have been steadily added to the network since the first days of RIPE Atlas, 96 of them being connected since 2010---they have been generating invaluable longitudinal data for the last 14 years. The great majority of active anchors have joined the system since 2018, which coincides with the introduction of virtual machine anchors. The oldest one has been running since 2013. We note, however, that the first connection dates do not necessarily signify that devices have been constantly connected ever since. Figure~\ref{fig:probe-total-uptime} additionally plots the total uptimes of connected probes and anchors measured in days. We can see that these findings are generally consistent with the expected uptimes based on the first connection dates. For example, the few probes first connected in 2010 have a total uptime of over 4.8~K days (over 13 years).

The consistent availability of RIPE Atlas probes and anchors highlights the network's reliability and stability. Probes that have remained connected for extended periods provide essential longitudinal data crucial for historical analysis. The consistency between their initial connection dates and ongoing uptime indicates near-continuous operation, ensuring reliable and uninterrupted data collection, essential for comprehensive insights into global Internet dynamics.

\begin{figure}[t]
    \begin{minipage}[b]{0.45\columnwidth}
        \includegraphics[width=\textwidth]{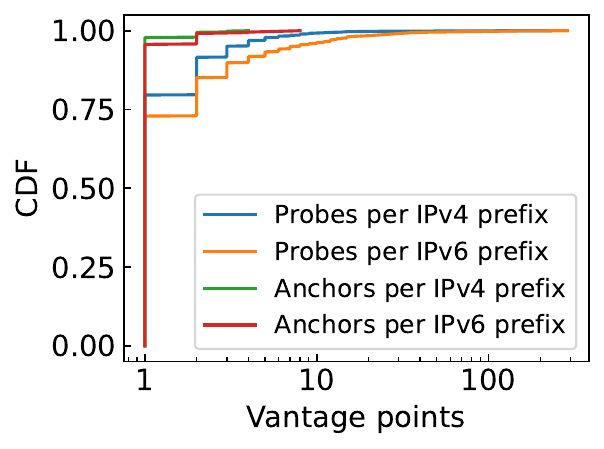}
        \caption{Number of RIPE Atlas probes and anchors connected to the platform on Feb 21, 2024, per IP prefix.}
        \Description{Number of RIPE Atlas probes and anchors connected to the platform on Feb 21, 2024, per IP prefix.}
        \label{fig:probes-per-prefix}
    \end{minipage}
    \hfill   
    \begin{minipage}[b]{0.45\columnwidth}
        \includegraphics[width=\textwidth]{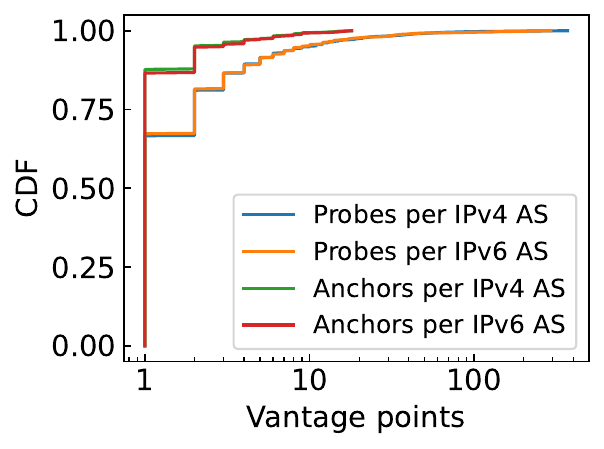}
        \caption{Number of RIPE Atlas probes and anchors connected to the platform on February 21, 2024, per AS.}
        \Description{Number of RIPE Atlas probes and anchors connected to the platform on February 21, 2024, per AS.}
    \label{fig:probes-per-asn}
    \end{minipage} 
\end{figure}

\subsubsection{IP Connectivity.} More than 12\% of connected probes are private, thus not revealing any information about their underlying IP connectivity. Focusing on public probes only, we see 5.6~K (46.5\%) of them being dual-stack, while the remaining 4.8~K (40\%) and 146 (1.2\%) are IPv4-only and IPv6-only, respectively. Anchors show a substantially higher support of IPv6, with 744 (92\%) being fully dual-stack. To further understand the distribution of these measurement vantage points, we compute the number of anchors and probes per IP prefix and AS. Figures~\ref{fig:probes-per-prefix}~and~\ref{fig:probes-per-asn} show the cumulative distributions for IPv4 and IPv6 address spaces. Overall, most IP prefixes and ASes with RIPE Atlas measurement vantage points contain exactly one probe or anchor, regardless of the address family. Anchors tend to be better distributed than probes, with the best coverage attained among IPv4 prefixes, for which 97.84\% contain only one anchor. In all other cases, each prefix contains at most 8 anchors. When aggregating the numbers at the AS level, we see some decrease in the diversity: the ratio of ASes with one anchor is 87-88\%, and at most, one probe is hosted in 67\% of ASes. The long tail also shows some ASes with a high concentration of probes, mostly Deutsche Telekom, Comcast Cable Communications, Free, and Orange. As they are major telecommunications operators in their corresponding countries, we hypothesize that probes may be hosted in the home networks of their owners.

The prevalence of private probes among RIPE Atlas measurements limits visibility into their IP connectivity, potentially skewing the analysis at the network level. The analysis of public probes reveals a notable dual-stack support among anchors compared to probes, highlighting varying network capabilities that researchers must consider when performing measurements across IPv4 and IPv6 address spaces.

\begin{figure}[t]
    \centering
    \includegraphics[width=\columnwidth]{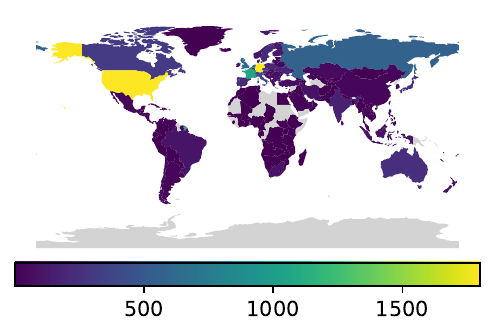}
    \caption{Number of connected RIPE Atlas probes and anchors per country in February 21, 2024.}
    \Description{Number of connected RIPE Atlas probes and anchors per country in February 21, 2024.}
    \label{fig-map}
\end{figure}

\subsubsection{Geographical Distribution.} We finally assess the geographical distribution of RIPE Atlas anchors and probes connected to the system as of February 21, 2024. Figure~\ref{fig-map} plots the number of probes and anchors per country. Overall, the platform has measurement devices in 178 countries worldwide. Germany and the United States host substantially more vantage points than any other country, both accounting for 28\% of probes and anchors. At the other extreme, 32 countries only have one probe or anchor. Generally, there is a strong bias towards Europe and North America, while RIPE Atlas remains underrepresented in other parts of the world.

\paragraph{\textbf{Key takeaways.}} \textit{Despite being under-represented in certain regions, RIPE Atlas boasts a wide coverage. Probes and anchors have been steadily joining the measurement network since 2010, serving as highly available and reliable baselines for analyzing the Internet. Their increasing support of both IP address families makes them suitable for any experiments requiring dual-stack support.}

\subsection{Measurements}

The 12.9~K active probes and anchors participated in 50,885 RIPE Atlas measurements, generating more than 1.3 billion individual results and 1.1~TB of raw data. Tables~\ref{tab:measurements-anchoring},\ref{tab:measurements-built-in},\ref{tab:measurements-user-defined} detail these statistics per anchoring, built-in, and user-defined measurements.

\subsubsection{Measurement Types.} The highest number of ongoing measurements (39~K or 76.7\%) were user-defined, most of them being pings and traceroutes. Pings are one of the most convenient ways to test the reachability of a particular host at a particular instant, thus being extremely popular among network operators. Overall, platform users launched all types of measurements available to them. Anchors contribute to the second largest chunk of ongoing measurements, but we recall that i) only anchors are targets of them, and ii) three types of measurements are performed (ping, traceroute, and HTTP). The remaining 251 measurements are built-in and are, therefore, preconfigured to run on each connected probe. More than half of them are DNS \texttt{CHAOS}-class \texttt{TXT} queries targeting all DNS root servers. We can also note that two types of measurements---specifically, probe (dis)connection events and traffic---are only applicable to built-ins. The latter is an opt-in feature that probe hosts can choose to collect statistics on the traffic received by the probe, for example, the number of received bytes or packets.

\subsubsection{Measurement Results.} One RIPE Atlas measurement may generate zero or more results, depending on the number of participating probes and periodicity. Out of 50.9~K ongoing measurements, 2.3~K did not generate a single result. In particular, 340 of them did not have a single participating probe, possibly due to the inability of the platform to accommodate the probe request, which may happen if a long-running measurement was defined with a set of probes not connected to the RIPE Atlas network anymore. Other cases included measurements launched and executed before the 24-hour analysis window but marked as active for a short period of time on February 21, 2024. We might have encountered active recurring measurements scheduled to run less frequently than every 24 hours, thus giving no results during the chosen day. Other 5.8~K measurements generated exactly one result object, meaning that they were either one-offs or recurring with the frequency of no more than once every 86.4~K seconds. In either case, one probe per measurement participated. 

A great majority of ongoing measurements generated two or more results, which becomes apparent when considering the average ratio of measurement results to measurements---over 26~K to 1. Due to the full mesh between all the anchors, they lead in the number and size of generated results despite being the most limited chunk of measurements in terms of sources, destinations, and types. Built-ins come second with over 283 M results and 182 GB of raw data. Even though only 251 measurements generated them, we recall that they are frequent (running up to every 4 minutes) and scheduled on all the connected probes. User-defined measurements generated the least amount of data.

\begin{table}[t]
    \caption{All the anchoring measurements executed, results (in millions), and the data generated (in gigabytes) on February 21, 2024. 
    \label{tab:measurements-anchoring}}
    \centering
    \scriptsize
    \setlength{\tabcolsep}{7pt}
    \begin{tabular}{lccc}
        \toprule
            \multirow{2}{*}{\textbf{Type}} & \multicolumn{3}{c}{\textbf{Anchoring}} \\
            \cmidrule(lr){2-4}
            & \textbf{Measurements} & \textbf{Results (M)} & \textbf{Size (GB)}\\
        \midrule
            Ping & 3,868 & 648.15 & 287.46 \\
            Traceroute & 3,841 & 172.08 & 419.26 \\
            DNS & - & - & - \\
            HTTP & 3,881 & 86.37 & 30.91 \\
            Probe (dis)connection  & - & - & - \\
            TLS  & - & - & - \\
            NTP & - & - & - \\
            Traffic & - & - & - \\
        \midrule
            \textbf{Total:} & 11,590 (22.8\%) & 906.6 (67.5\%) & 737.63 (68.7\%) \\
        \bottomrule
    \end{tabular}
\end{table}

\begin{table}[t]
    \caption{All the built-in measurements executed, results (in millions), and the data generated (in gigabytes) on February 21, 2024. 
    \label{tab:measurements-built-in}}
    \centering
    \scriptsize
    \setlength{\tabcolsep}{7pt}
    \begin{tabular}{lccc}
        \toprule
            \multirow{2}{*}{\textbf{Type}} & \multicolumn{3}{c}{\textbf{Built-in}} \\
            \cmidrule(lr){2-4}
            & \textbf{Measurements} & \textbf{Results (M)} & \textbf{Size (GB)}\\
        \midrule
            Ping & 38 & 127.77 & 54.63 \\
            Traceroute & 44 & 21.51 & 50.79 \\
            DNS & 158 & 126.91 & 74.95 \\
            HTTP & 4 & 0.47 & 0.23 \\
            Probe (dis)connection  & 2 & 6.14 & 1.1 \\
            TLS  & 4 & 0.07 & 0.25 \\
            NTP & - & - & - \\
            Traffic & 1 & 0.68 & 1.0 \\
        \midrule
            \textbf{Total:} & 251 (0.5\%) & 283.55 (21.1\%) & 182.95 (17.0\%) \\
        \bottomrule
    \end{tabular}
\end{table}

\subsubsection{Creation Times.} 
\begin{figure}[t]
    \centering
    \begin{minipage}[b]{0.44\columnwidth}
        \includegraphics[width=\textwidth]{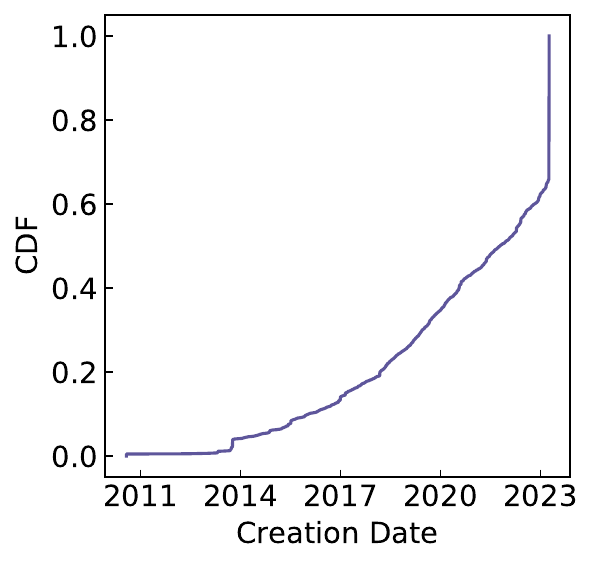}
        \caption{Dates when measurements active on February 21, 2024 were first created.}
        \Description{Dates when measurements active on February 21, 2024 were first created.}
        \label{fig:msm-creation-date}
    \end{minipage}
    \hfill   
    \begin{minipage}[b]{0.45\columnwidth}
        \includegraphics[width=\textwidth]{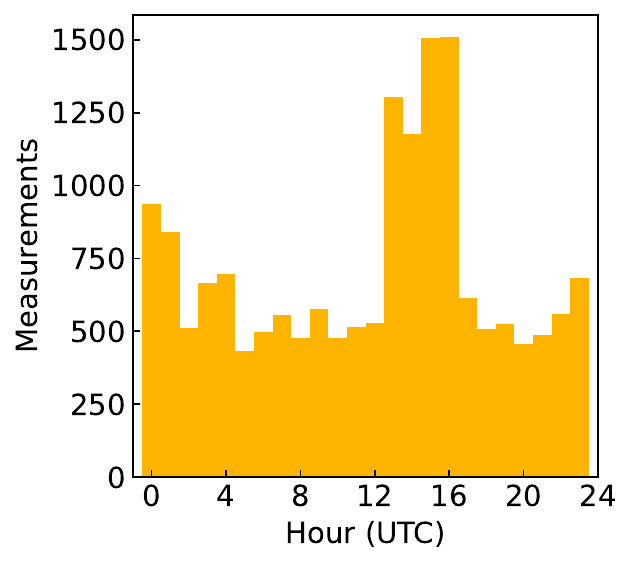}
        \caption{Creation times of measurements that were launched on February 21, 2024.}
        \Description{Creation times of measurements that were launched on February 21, 2024.}
        \label{fig:msm-creation-hour}
    \end{minipage}
\end{figure}

We next analyze whether the 50.8~K measurements observed on February 21, 2024, were first launched on the same day or they had been running for some time. Figure~\ref{fig:msm-creation-date} presents the distribution of measurement creation dates, showing that 240 built-ins have been running since the first days of RIPE Atlas in 2010, collecting results from all the connected probes ever since. A significantly larger chunk of ongoing measurements (17~K or 33.5\%) were first executed during the 24-hour analysis window. Moreover, most of them were also stopped the same day, suggesting that an important part of daily active measurements is either one-off or short-term. Figure~\ref{fig:msm-creation-hour} additionally plots the number of measurements created per hour (UTC) on February 21, 2024. There was a burst of 3,276 traceroute measurements between approximately 13:00~UTC and 16:00~UTC, seemingly belonging to the same campaign, as all the descriptions started with ``Traceroute to''. They concern 295 domain names with up to 15 measurements per domain, however without any clear pattern or any domain-specific selection strategy.

\begin{table}[t]
    \caption{All the user-defined measurements executed, results (in millions), and the data generated (in gigabytes) on February 21, 2024. 
    \label{tab:measurements-user-defined}}
    \scriptsize
    \centering
    \setlength{\tabcolsep}{7pt}
    \begin{tabular}{lccc}
        \toprule
            \multirow{2}{*}{\textbf{Type}} & \multicolumn{3}{c}{\textbf{User-defined}} \\
            \cmidrule(lr){2-4}
            & \textbf{Measurements} & \textbf{Results (M)} & \textbf{Size (GB)} \\
        \midrule
            Ping & 19,225 & 33.81 & 15.21 \\
            Traceroute & 10,831 & 31.96 & 59.95 \\
            DNS & 7,898 & 83.15 & 65.9 \\
            HTTP & 3,881 & 86.37 & 30.91 \\
            Probe (dis)connection & - & - & - \\
            TLS  & 670 & 2.98 & 11.86 \\
            NTP & 166 & 0.2 & 0.13 \\
            Traffic & - & - & - \\
        \midrule
            \textbf{Total:} & 39,044 (76.7\%) & 153.55 (11.4\%) & 153.6 (14.3\%) \\
        \bottomrule
    \end{tabular}
\end{table}

\subsubsection{Measurement Origins.} We wanted to know \textit{who} launched all 39~K user-defined measurements, keeping in mind that this information is not publicly available and there is no mapping between measurement IDs and end users. Instead, we refer to the description field available for each measurement. If not using the default string provided by the system, it may give a good indication of who is behind a particular measurement. It turns out that some of the user-defined measurements were in fact launched to back up certain RIPE NCC services. For example, DNSMON~\cite{dnsmon-visualising} is behind 4,435 executed measurements to assess the quality of service provided by root servers and selected top-level domains (TLDs). DomainMON~\cite{domainmon} (1,684 measurements) is a similar tool that one can set up to monitor nameservers of its own domains. We also see 1,527 measurements triggered by RIPE IPmap~\cite{ipmap}, an active geolocation API relying on RIPE Atlas ping measurements under the hood. Focusing on external measurements, there are 598 pings, traceroutes, and DNS queries with the description of ``HE Network Tools Site''. A closer look reveals that they are launched from \url{https://bgp.he.net/traceroute/} -- a measurement toolkit run by Hurricane Electric~\cite{aggregators}. We then see a number of ping measurement campaigns targeting the infrastructure of DNS operators, ISPs, content delivery networks (CDNs), and one regional Internet registry, among others.

\begin{figure}[t]
    \begin{minipage}[b]{0.45\columnwidth}
        \includegraphics[width=\textwidth]{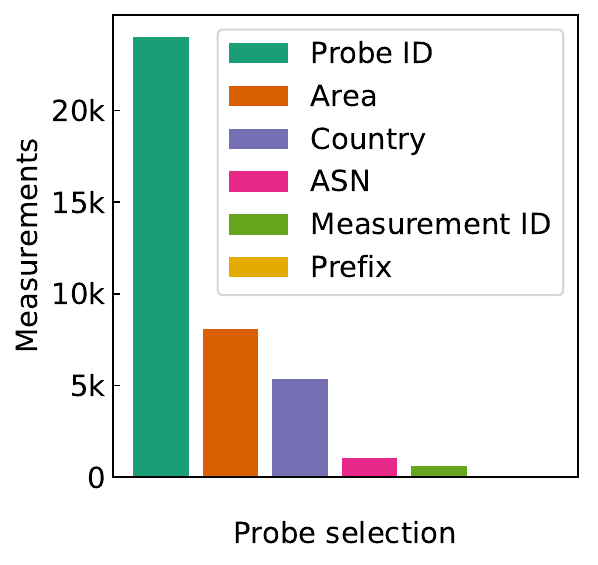}
        \caption{Criteria used by RIPE Atlas users to choose probes for measurements.}
        \Description{Criteria used by RIPE Atlas users to choose probes for measurements.}
        \label{fig:probe-selection}
    \end{minipage}
    \hfill   
    \begin{minipage}[b]{0.44\columnwidth}
        \includegraphics[width=\textwidth]{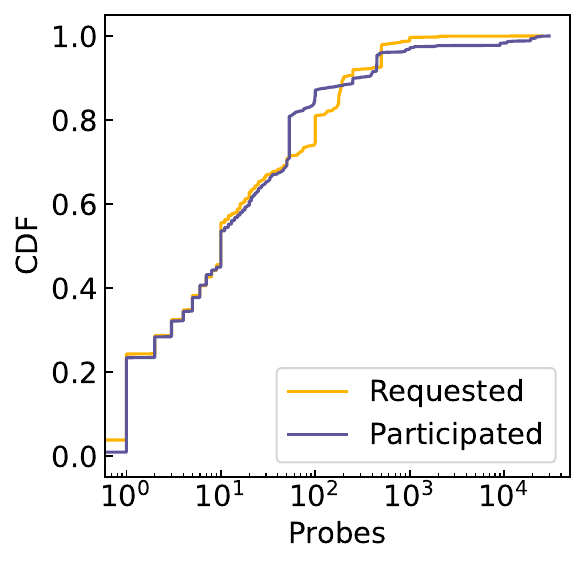}
        \caption{Distribution of the number of requested and actually participating probes.}
        \Description{Distribution of the number of requested and actually participating probes.}
        \label{fig:probe-requests}
    \end{minipage}
\end{figure}

\subsubsection{Measurement Definitions.} Before launching a measurement, a RIPE Atlas user needs to choose one or more probes by supplying either i) a probe ID, ii) a geographic area, iii) a country, iv) an AS, v) an IP prefix, or vi) an existing measurement ID to reuse the same list of participants. Focusing on 39~K user-defined measurements, we analyze the participant selection criteria requested by end users. Figure~\ref{fig:probe-selection} plots the distribution of participation request types. Probe IDs are by far the most common way to choose measurement sources, although the two geographical selectors are also largely used. The chosen area was rarely a particular part of the world but rather ``worldwide'' and the US, Germany, and France were the most requested source countries. 

Despite the possibility of freely choosing participating probes, the probe may be disconnected or too busy to accommodate more measurements at the time. Figure~\ref{fig:probe-requests} plots the distribution of the number of selected probes and the number of actually participating probes. Few user-defined measurements (337) did not have a single participating probe, most of them being requested with probe IDs. We further verified that a great majority of requested probes were not connected on the day of the measurement, therefore being unable to process the requests. Almost 20~K measurements were executed with exactly the same sources as defined and the remaining 19~K has a subset of desired participants.

\subsubsection{Measurement Targets.} Finally, we assess the distribution of measurement targets, i.e., the systems being tested. Pings were sent towards 8~K unique hosts, the top 5 being one DNS provider from China and 4 TikTok CDN servers. They represented 20\% of all daily ping measurements, as they participated in 4.9~K measurements in total. They are also the most common measurement destinations across all types of measurements. Traceroutes were distributed towards 6.5~K destinations, without having any significant outliers. 

DNS measurements differ from other types as their targets are domain names rather than Internet hosts. However, the two pseudo domain names (namely \texttt{hostname.bind} and \texttt{version.bind}) leading the ranking are not globally resolvable, and the responses are, therefore, to be provided by the destination DNS servers. Despite HTTP measurements only targeting the RIPE Atlas infrastructure, we found 7 hosts outside the list of normally allowed destinations, including two regional Internet registries, two connectivity monitoring domains from a leading operating systems vendor, one lifestyle blog, one data center, and one domain belonging to a search engine operator. TLS and NTP measurements targeted 560 and 137 unique hosts, respectively, almost always being unique per measurement. Interestingly, 80 user-defined measurements have overlapping types and targets with some of the built-ins. While they generated a negligible number of results (4.9~M) with respect to all produced in a single day, we argue that built-ins could have been used instead.

\paragraph{\textbf{Key takeaways.}} \textit{A great majority of ongoing measurements in a day are user-defined, but it is the anchoring meshes and built-ins that generate almost 89\% of daily results. Overall, more than 39~K measurements were launched by end users in a day, mostly running on particular requested probes. However, some of them are not standalone and back up certain RIPE NCC services as well as external measurement platforms, a grey area in the usage of RIPE Atlas.}

\section{Measurement Use Cases\label{sec:day-in-the-life}}

The previously described dataset contains more than 1.3 billion individual measurement results. In this section, we present a series of exploratory use cases observed on February 21, 2024, illustrating how RIPE Atlas data can uncover emerging patterns and open new avenues for research. 

\subsection{Anchoring Measurements\label{subsec-anchoring-measurements}}

The majority of results generated by the RIPE Atlas platform in a single day come from anchoring measurements. Below, we describe how to use them to analyze traceroute symmetry. 

\subsubsection{Traceroute Symmetry.} The full mesh traceroute measurements performed between all the anchors offer a rare opportunity to analyze the symmetry of paths on the Internet, as the anchor measurements provide a large set of bidirectional traceroutes in close temporal proximity. The asymmetry of Internet paths has gained attention in the past~\cite{YihuaHe2005,DeVries2015,reverse-traceroute} and was studied both via the analysis of BGP data and active measurements using traceroute. Understanding path symmetry is useful for troubleshooting cases like routing problems that occur on the reverse path or path asymmetry that leads to differences in latency affecting latency-sensitive applications like video conferencing.

While there are studies that employed RIPE Atlas for their traceroute measurements~\cite{DeVries2015}, they only scheduled their own measurements. Thus, they faced issues with the daily credit limit and had to make compromises for their measurement configuration: Instead of a full mesh measurement, which would only include 112 probes due to the credit limit, they opted to perform pairwise measurements between 4~K random probes, resulting in 2~K pairs. While anchors were not widely deployed back then, we can now leverage the data generated by 810 anchors without having to schedule a single measurement on our own.

\begin{figure}[t]
    \centering
    \includegraphics[width=0.75\linewidth]{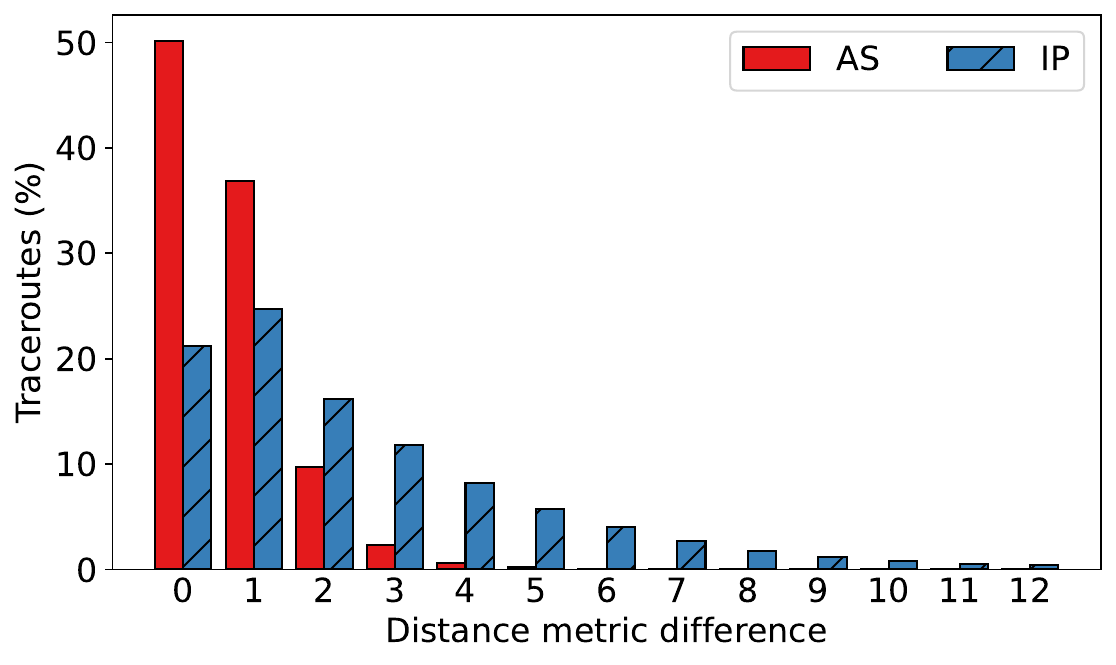}
    \caption{Difference of traceroute lengths computed based on anchoring measurements.}
    \Description{Difference of traceroute lengths computed based on anchoring measurements.}
    \label{fig:traceroute-symmetry}
\end{figure}

We find that from the 810 active anchors, only 770 generated valid traceroutes during our chosen day. In total, the measurements result in 292.7~K anchor pairs and almost 28 M traceroutes suitable for analysis. A detailed analysis of symmetry is beyond the scope of this paper, but we provide a preliminary estimation by comparing the length of traceroutes in both directions. If the length is different, the traceroutes cannot be symmetric. We use two length metrics: i) the number of IP hops and ii) the number of unique ASes traversed by each traceroute. The analysis reveals that only 21\% of traceroutes could be symmetric, having the same length in terms of IP hops in both directions (see Figure~\ref{fig:traceroute-symmetry}). The majority of traceroutes is longer by one hop or more in one direction, making symmetry impossible. However, considering the length at the AS granularity, around 50\% of traceroutes traverse the same number of ASes in both directions and are potentially symmetric. 

\subsection{Built-in Measurements\label{subsec-built-measurements}}

Built-in measurements, which have been running on every connected RIPE Atlas host since 2010, offer a wealth of long-term data for exploration. In this section, we highlight practical use cases that illustrate the potential of these measurements, revealing insights into network behavior.

\subsubsection{Manipulation of the Popular Domain Resolution.} One of the built-in measurements queries local resolvers to obtain \texttt{A} and \texttt{TXT} resource records of 50 popular domain names. Previous work already relied on user-defined RIPE Atlas measurements to study DNS resolution manipulation~\cite{intercept-and-inject}. Below, we show how one can detect response injection using built-ins only. 

We first consulted the Open Observatory of Network Interference (OONI)~\cite{ooni} project and identified that 17 out of 50 queried domains show strong indications of blocking in at least one country. For each of them, we parsed the responses to \texttt{A} requests and extracted the returned IPv4 addresses. We cannot, however, create one-to-one mappings between the domains and \texttt{A} records as the DNS load balancing systems return different IP addresses depending on the location and the time of the query. Therefore, we estimate a range of allowed responses by first resolving each domain directly at \url{https://dns.google} (Google Public Resolver) and then mapping responses to ASes~\cite{pyasn}. We consider a response as valid if it falls within the range of ASes seen in responses from Google Public Resolver. While this non-strict approach may potentially underestimate the amount of blocking (i.e., IP belonging to an observed AS is not, in fact, a correct address for a particular domain), it does highlight the cases for which the answer is incorrect. We also note that other strategies exist to restrict access to a particular domain at the DNS level, for example, returning the \texttt{NXDOMAIN} or \texttt{SERVFAIL} response codes. We argue, however, that the codes are indistinguishable from intermittent failures, while response injection is a strong indication of deliberate manipulation.

The highest number of probes received injected responses for two Meta services: \texttt{facebook.com} and \texttt{instagram.com}. They were blocked on at least 69\% of probes from China and Iran, mostly receiving globally routable IPs that belong to well-known organizations such as Twitter, Dropbox, or Amazon. Such a behavior is consistent with the effects of the Great Firewall of China~\cite{how-great}. The two countries restricted access to other entertainment domains such as \texttt{youtube.com}, \texttt{reddit.com}, \texttt{twitter.com}, and \texttt{netflix.com}, as it can be seen on more than half of the countries' probes. The third most blocked domain was \texttt{vk.com}, a Russian social media network with 200 affected RIPE Atlas probes, 60 of which were from Iran alone (68\% of the country probes). The two countries formally blocking this domain name, namely Ukraine and Latvia, have 18\% and 37\% of probes receiving injected responses, respectively. Some of the other domain names triggering injected responses are \texttt{google.com}, \texttt{google.co.in}, and \texttt{linkedin.com}---they were blocked on the majority of probes located in China. Similarly, \texttt{t.co} and \texttt{wordpress.com} were not reachable for the majority of RIPE Atlas probes located in Iran.

Overall, we demonstrate how RIPE Atlas built-in measurements can be used to study DNS interception, manipulation, and censorship of popular domains, complementing some of the existing tools like OONI. They can give insight into blocking strategies used on a per-probe basis, e.g., injecting globally routable or private IP addresses. When interpreting results, it is crucial to take into account the number of probes per country to avoid false inference of nationwide blocking. 

\subsubsection{Unallocated 240/4 in the Wild.} 
The IPv4 space has many unallocated or reserved addresses with the largest block of 240/4. Two proposals suggested repurposing 240/4 into unicast~\cite{proposal-240-1} or private~\cite{proposal-240-2} address spaces. Neither proposal was adopted though, as critics argued that these additional addresses would be quickly exhausted, emphasizing the need to transition to IPv6. Previous work revealed that some organizations internally use the IP address space allocated to others but never announced on the Internet~\cite{squatters}. Below, we analyze all the built-in traceroutes collected on 21 February 2024 to see whether the same phenomenon is observed for unallocated address ranges.

Overall, we identified 1.7~M traceroutes containing 240/4 addresses. Table~\ref{tab:traceroutes} further lists all the autonomous systems hosting the probes from which the traceroutes originate. Similarly to previous reports~\cite{240-4}, the majority of them (92.7\%) originate from two Amazon ASes  -- AS16509 and AS14618. We saw unallocated IPs in the first and second hops, indicating the constant internal use of this prefix by Amazon. Similar behavior was observed at traceroutes originating from Telefonica Spain (AS3352), where hops subsequent to 240/4 remained inside the same AS. Despite the previous hops timing out, we can still conclude that Telefonica uses the 240/4 address space internally. Traceroutes originating from the remaining autonomous systems (AS8728, AS577, AS398721) seem to merely traverse the Amazon AS that makes 240/4 appear in some of the transit hops.

The use of the unallocated/squatted address space is strongly discouraged, as it might render parts of the Internet unavailable should the concerned IP addresses become globally routable. RIPE Atlas measurements reveal that at least two organizations use the unallocated 240/4 address block internally. Another key finding is that all the networks we observed in the traceroutes do not have proper filtering in place, as packets with addresses in the 240/4 prefix should be dropped.

\begin{table}[t!]
    \centering
    \scriptsize
    \caption{The number of traceroutes containing 240/4 hops per originating autonomous systems.}
    \label{tab:traceroutes}
    \begin{tabular}{clcc}
        \toprule
            \textbf{AS Number} & \textbf{Organization} & \textbf{Country} & \textbf{Number of Traceroutes} \\
        \midrule
            16509 & Amazon.com, Inc. & US & 1,601,758 \\
            14618 & Amazon.com, Inc. & US & 106,888 \\
            3352 & TELEFONICA DE ESPANA S.A.U. & ES & 10,128 \\
            8728 & AS INFONET & EE & 8,502 \\
            577 & Bell Canada & CA & 458 \\
            398721 & Cogeco Connexion inc & CA & 30 \\
        \bottomrule
    \end{tabular}
\end{table}

\subsubsection{Unspecified ::/128 as a Source Address.} Some of the IP addresses are called special use and are not intended to be assigned to hosts. One such example is ::/128, called \textit{unspecified} and prohibited from being used neither as a source nor as a destination~\cite{rfc4291}. Nevertheless, it was reported that some of the traceroutes collected by RIPE Atlas contain hops with ::/128 source addresses~\cite{128}. We verify whether the same behavior persisted on the day of the analysis. 

On February 21, 2024, RIPE Atlas gathered 95~M IPv6 traceroute results. We found that 334~K out of them (0.35\%) include at least one hop with the ::/128 source address. All traceroutes originated from 161 probes, including 1 probe from NeuStyle (AS4508) responsible for 256~K (76\%) cases observed. This probe has ::/128 consistently appearing as the first hop. In the remaining cases, we mapped the IP addresses of previous and subsequent hops to autonomous system numbers. We reveal several traces with AS24186 (RailTel Corporation of India Ltd.) surrounding the ::/128 hop. Previously, it was AS9198 (JSC Kazakhtelecom) that was seen before and after the ::/128 hop. Although we do not see the same AS in our dataset, we observed some instances where the hop preceding the address belonged to either AS12389 or AS20485, and the subsequent hop was in AS9198 indicating that the phenomenon is still persistent in this network. 

Overall, our analysis reveals that some ASes violate the requirements of the RFC~4291 and route packets with unspecified source addresses. This behavior also reveals the lack of Source Address Validation (SAV) \cite{sav-sp} in these networks.

\subsection{User-Defined Measurements\label{subsec-user-defined}}

The two previously discussed groups of measurements are well-known and have been defined by the RIPE Atlas team. Below, we look at some of the user-defined measurements to understand what they may reveal.

\subsubsection{TikTok CDN}

We observed one noticeable user-defined measurement that involved periodic pings to four TikTok CDN servers. They account for 25\% (4,863) of user-defined ping measurements but generated only 26.5~K results (less than 0.1\%) because of the way the measurements are set up: the four servers are targeted from 50 countries in hourly cycles. However, for reasons unknown to us, the creators of the campaign chose to create individual one-off measurements instead of one long-running periodic measurement per country. Our best guess is that the creators use a custom scheduling setup, supported by the fact that even though measurements are run per country (indicated by a ``Ping from \emph{country}'' description), they do not use the country selection provided by Atlas, but request individual probes by ID instead. Thus, the measurements also significantly contribute to the probe source distribution shown in Figure~\ref{fig:probe-selection}. We speculate that---for the most part---using periodic measurements
instead of one-off wold have been more economical for this purpose.

\begin{figure}[t]
    \centering
    \includegraphics[width=1\linewidth]{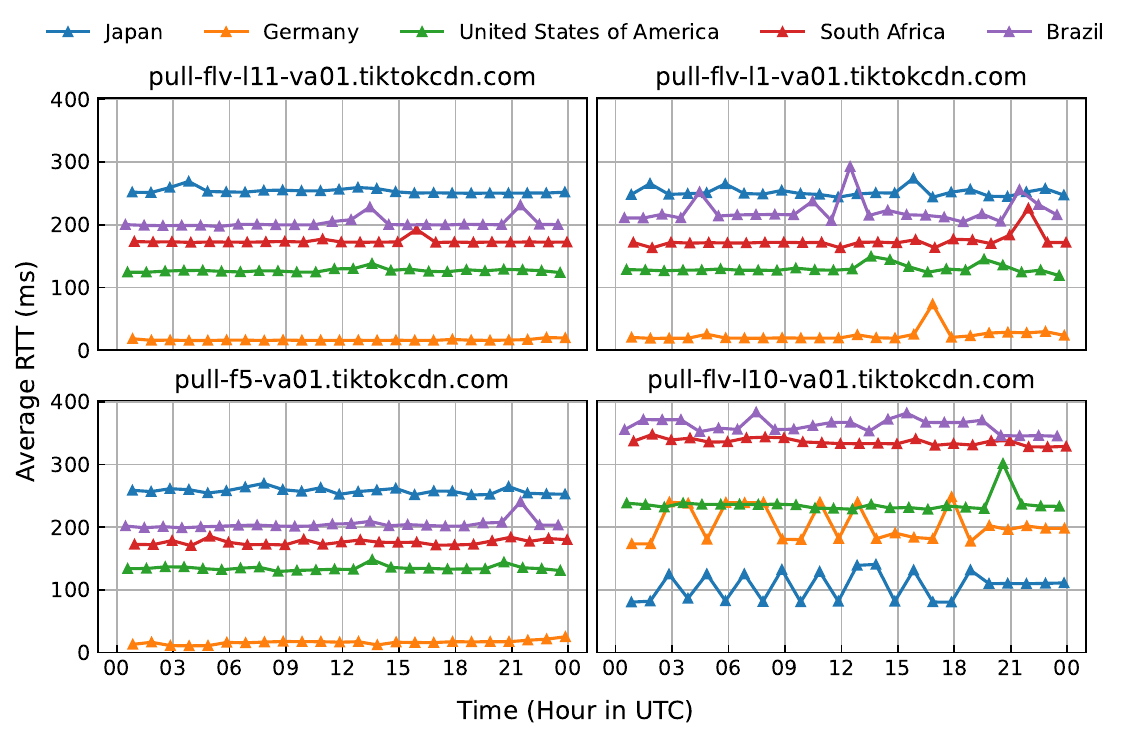}
    \caption{Average RTT from Japan, Germany, United States, South Africa, and Brazil to four TikTok CDN servers.}
    \Description{Average RTT from Japan, Germany, United States, South Africa, and Brazil to four TikTok CDN servers.}
    \label{fig:tiktok}
\end{figure}

A rough search reveals that this campaign has been running since June 2023 and created 2.5~M measurements since then. The frequency (hourly) and the type of measurement (ping) make this dataset useful to track long-term changes in RTT from different countries to the CDN servers. Figure~\ref{fig:tiktok} presents the average RTT observed from five countries during our analyzed day. It is apparent that the connectivity was stable during this day and the server \texttt{flv-l10} seems to be located in the Asian region, since the average RTT from Japan is the lowest for this server, whereas the country order is stable for the other three targets. A deeper analysis of this dataset might reveal intermittent connectivity problems from certain countries or long-term changes in RTT.

\subsubsection{IXP Country and Regional Jedi Measurements.} RIPE Atlas team set up a number of traceroute meshes that aim at analyzing the connectivity between probes located inside the same country and region. These measurements offer critical information on the paths taken, including out-of-country detours and the IXPs the traffic traverses.

Country-level meshes are executed on the first day of each month. These are regular and up-to-date measurements that can be used to analyze how traffic patterns differ between countries. For example, they reveal the presence of 17 IXPs operating in Germany and 14 IXPs in the Netherlands. In the United States, they identify 66 unique peering LANs in probe paths, associated with 45 unique IXPs. Conversely, no IXPs are identified in Uzbekistan, China, and Iran. However, the traceroutes between ASes in China and Uzbekistan predominantly pass through the incumbent providers Uzbektelecom and the China Backbone network for the majority of the traceroutes.

Regional meshes have been collecting traceroutes in Southeast Europe, Central Asia, Latin America, and the Middle East on the 15th of every month since January 2024. These measurements originate from several countries within each region, highlighting the involvement of intermediary countries and ASes. We therefore focus on the data collected in the Central Asia region, including Uzbekistan, Kazakhstan, Iran, Kyrgyzstan, and Tajikistan. Overall, we see that Central Asian countries heavily rely on intermediaries outside the region to communicate with other hosts inside. Russia is by far the country that received the highest ratio of transit traffic, ranging from 42.4\% of traceroutes in Kazakhstan to as many as 72.6\% in Tajikistan and 80.7\% in Iran. The autonomous systems involved are PJSC Rostelecom (AS12389), PJSC ``Vimpelcom'' (AS3216), and Inetcom LLC (AS35598). Azerbaijan received the second highest amount of traffic, always transiting through Delta Telecom Ltd (AS29049) from Uzbekistan (19.2\% of traceroutes), Kazakhstan (24.2\%), and Iran (46.9\%). We do see a much smaller number of cases in which traffic goes to ASes registered in as far as Sweden, Italy, and the United States.

Based on the regional traceroute meshes collected by RIPE Atlas, Central Asia is highly dependent on several foreign countries and ASes to transit its traffic. Such a lack of connectivity options is generally expensive, not flexible, and may result in a single point of failure. The five aforementioned countries will benefit from consolidating efforts to build a more diverse internal interconnection.

\section{Guidelines for Using RIPE Atlas\label{sec-call}}

RIPE Atlas is a powerful tool that executes diverse measurements and generates over a billion results daily. As a measurement community reliant on its capabilities, we have a unique opportunity to optimize our use of its resources. Therefore, we offer guidelines for researchers to enhance their usage of RIPE Atlas, which we proposed to and extensively discussed with the RIPE Atlas team.

\subsection{Optimize the Infrastructure Usage}

RIPE Atlas operates with a set of safeguards to ensure that no single user or measurement jeopardizes the whole platform. 
Apart from adhering to internally enforced measurement constraints, researchers must consider the volume of generated data and avoid performing unnecessary measurements, especially for extended periods beyond those necessary to address their research questions.

One might feel compelled to launch a new measurement campaign as soon as the research question is clearly defined. However, we recommend stepping back and carefully considering whether any existing measurements can serve the purpose instead. The platform offers a variety of built-ins and used-defined measurements (documented in Measurement Bundles~\cite{bundles} repository), which already run on RIPE Atlas. Reusing existing data helps avoid redundant efforts. Each research paper using RIPE Atlas must include a statement addressing whether built-in measurements were considered to answer the research question and justify the need for any new measurements.

Other examples of suboptimal behaviors have already been provided~\cite{anti-patterns}, such as failing to use a single recurring measurement instead of multiple one-offs or polling for results more frequently than they are generated. Carefully reading the relevant documentation will help avoid such non-optimal behaviors.

\subsection{Consider Ethics}
In addition to relevant guidelines and best practices for the measurement community~\cite{considerations,zmap}, authors should carefully consider specific ethical implications related to the RIPE Atlas platform to minimize unforeseen consequences for all users.

If a new measurement is deemed necessary, researchers must assess its ethical aspect from multiple angles. First, the Terms and Conditions~\cite{terms} of the platform must serve as a starting point to understand the expectations from all participants. In particular, it states that probe owners agree that the installed devices are used for performing measurements and obtaining the results. Yet, one must ensure that no risk is put on the hosts. For example, avoid sending DNS requests to sensitive domain names that may be banned under certain jurisdictions such as related to gambling or adult content. We also recommend that researchers consult the RIPE community discussion on measurement ethics~\cite{ethics}.

\subsection{Encourage Reproducibility}

RIPE Atlas offers various mechanisms, such as tags and descriptions, to facilitate finding relevant measurements and foster reproducibility. We strongly advise researchers to assign unique tags and provide comprehensive descriptions for all new measurements. They, along with measurement IDs, should be referenced in scientific publications for easy data access. Furthermore, documenting new user-defined measurement campaigns in the Measurement Bundles repository is encouraged to inform others and promote the reuse of existing measurements whenever feasible.

\subsection{Explore the RIPE Atlas Data}

RIPE Atlas is often used to investigate specific research questions or validate hypotheses. However, given its extensive measurement dataset, there may be numerous events and insights beyond one's immediate focus. We encourage researchers to explore existing RIPE Atlas measurements thoroughly and uncover findings that could benefit the network community. We firmly believe that this rich dataset holds significant potential for discovering new insights and phenomena.

\subsection{Host a Probe}

Participants around the globe play a crucial role in expanding the global presence of RIPE Atlas by hosting probes---essential components of the measurement network. We strongly encourage researchers and network operators to consider applying to host either a hardware or a software probe, which can be installed and managed on their own infrastructure. Active involvement from users in underrepresented regions is particularly encouraged. For detailed instructions, interested researchers can visit the official RIPE Atlas web page.

\section{Ethics and Reproducibility\label{sec-ethics}}

Large-scale Internet measurements typically do not fall under the purview of Institutional Review Boards. Therefore, researchers must carefully weigh the advantages and disadvantages before initiating any new measurement campaign. To aid in this process, the community has developed a set of best practices to adhere to~\cite{zmap,considerations,menlo}.

This paper strictly adheres to the RIPE Atlas Terms and Conditions regarding data retrieval, analysis, and presentation. Specifically, all network configurations, measurements, and probes discussed are publicly accessible, as agreed upon by probe hosts in \S4.3 and \S4.2 of the terms. We solely use existing datasets to uncover various phenomena observed on the analysis day, ensuring an unbiased exploration. Recognizing the significant data volume, we carefully refer to relevant documentation to prevent overloading the RIPE Atlas infrastructure with our requests. 

\section{Conclusions\label{sec-conclusions}}

RIPE Atlas stands as one of the largest network measurement platforms worldwide, executing 50.9~K unique measurements and generating over 1.3 billion results daily. This unprecedented scale presents challenges in understanding its diverse user base, methodologies, and objectives. 

We examined a typical day in the life of RIPE Atlas to uncover usage patterns. The majority of daily measurements are user-defined, with nearly half being one-off tests. Anchors contribute significantly, generating almost 70\% of daily results, surpassing both built-in (21.1\%) and user-defined measurements (11.4\%). Pings, traceroutes, and DNS tests dominate, with HTTP, TLS, and NTP lagging behind. End users primarily perform one-off measurements and do not heavily burden the system.

Probes and anchors have steadily expanded the RIPE Atlas network since 2010, providing robust baselines for Internet analysis. Around half of the public probes support IP dual-stack, while 92\% of anchors are fully dual-stack. Despite uneven representation across regions, RIPE Atlas spans 178 countries and over 4,000 ASes.

Given the data volume, RIPE Atlas proves invaluable for diverse research and operational questions. With exclusively existing datasets, we have demonstrated how anchoring, built-in, and user-defined measurements can illuminate traceroute symmetry, address space usage, DNS censorship, and more.

Lastly, after extensive discussions with the RIPE Atlas team, we propose guidelines for researchers leveraging RIPE Atlas: careful network usage, full reproducibility, maximal reuse of existing measurement data, and rigorous ethical considerations.

\bibliographystyle{splncs04}
\bibliography{references}

@article{considerations,
    author = {Partridge, Craig and Allman, Mark},
    title = {{Ethical Considerations in Network Measurement Papers}},
    year = {2016},
    issue_date = {October 2016},
    publisher = {ACM},
    volume = {59},
    number = {10},
    journal = {Commun. ACM},
    month = {sep},

}

@inproceedings{zmap,
    author = {Durumeric, Zakir and Wustrow, Eric and Halderman, J. Alex},
    title = {{ZMap: Fast Internet-Wide Scanning and Its Security Applications}},
    year = {2013},
    booktitle = {USENIX Security},
    publisher = {USENIX Association},
}

@article{ripe-paper,
    author = {{RIPE NCC Staff}},
    year = {2015},
    month = {01},
    title = {{RIPE Atlas: A Global Internet Measurement Network}},
    volume = {18},
    journal = {The Internet Protocol Journal}
}

@inproceedings{interference,
    author = {Holterbach, Thomas and Pelsser, Cristel and Bush, Randy and Vanbever, Laurent},
    title = {{Quantifying Interference between Measurements on the RIPE Atlas Platform}},
    year = {2015},
    publisher = {ACM},
    booktitle = {IMC},

}

@article{lessons-learned,
    author = {Bajpai, Vaibhav and Eravuchira, Steffie Jacob and Sch\"{o}nw\"{a}lder, J\"{u}rgen},
    title = {{Lessons Learned From Using the RIPE Atlas Platform for Measurement Research}},
    year = {2015},
    publisher = {ACM},
    journal = {SIGCOMM Comput. Commun. Rev.},
    month = {jul},
    volume = {45},
    number = {3},
}

@inproceedings{intercept-and-inject,
    author={Nosyk, Yevheniya and Lone, Qasim and Zhauniarovich, Yury and Ga{\~{n}}{\'a}n, Carlos H. and Aben, Emile and Moura, Giovane C. M. and Tajalizadehkhoob, Samaneh and Duda, Andrzej and Korczy{\'{n}}ski, Maciej},
    title={{Intercept and Inject: DNS Response Manipulation in the Wild}},
    booktitle={{PAM}},
    year={2023},
    publisher={Springer Nature Switzerland},
    
    
}

@inproceedings{internet-top-lists,
    author = {Scheitle, Quirin and Hohlfeld, Oliver and Gamba, Julien and Jelten, Jonas and Zimmermann, Torsten and Strowes, Stephen D. and Vallina-Rodriguez, Narseo},
    title = {{A Long Way to the Top: Significance, Structure, and Stability of Internet Top Lists}},
    year = {2018},
    publisher = {ACM},
    booktitle = {IMC},

}

@inproceedings{old-but-gold,
    author={Moura, Giovane C. M. and Heidemann, John and Hardaker, Wes and Charnsethikul, Pithayuth and Bulten, Jeroen and Ceron, Jo{\~a}o M. and Hesselman, Cristian},
    title={{Old but Gold: Prospecting TCP to Engineer and Live Monitor DNS Anycast}},
    booktitle={PAM},
    year={2022},
    publisher={Springer International Publishing},
    
    
}

@inproceedings{manycast2,
    author = {Sommese, Raffaele and Bertholdo, Leandro and Akiwate, Gautam and Jonker, Mattijs and van Rijswijk-Deij, Roland and Dainotti, Alberto and Claffy, KC and Sperotto, Anna},
    title = {{MAnycast2: Using Anycast to Measure Anycast}},
    year = {2020},
    publisher = {ACM},
    booktitle = {IMC},

}

@inproceedings{taming,
    author = {McQuistin, Stephen and Uppu, Sree Priyanka and Flores, Marcel},
    title = {{Taming Anycast in the Wild Internet}},
    year = {2019},
    publisher = {ACM},
    booktitle = {IMC},

}

@inproceedings{regional-anycast,
    author = {Zhou, Minyuan and Zhang, Xiao and Hao, Shuai and Yang, Xiaowei and Zheng, Jiaqi and Chen, Guihai and Dou, Wanchun},
    title = {{Regional IP Anycast: Deployments, Performance, and Potentials}},
    year = {2023},
    publisher = {ACM},
    booktitle = {SIGCOMM},

}

@inproceedings{anycast-context,
    author = {Koch, Thomas and Katz-Bassett, Ethan and Heidemann, John and Calder, Matt and Ardi, Calvin and Li, Ke},
    title = {{Anycast in Context: a Tale of Two Systems}},
    year = {2021},
    publisher = {ACM},
    booktitle = {SIGCOMM},

}

@inproceedings{anycast-agility,
    author = {A S M Rizvi and Leandro Bertholdo and Jo{\~a}o Ceron and John Heidemann},
    title = {{Anycast Agility: Network Playbooks to Fight DDoS}},
    booktitle = {USENIX Security},
    year = {2022},
    publisher = {USENIX Association},
    month = aug
}

@inproceedings{replication-geolocation,
    author = {Darwich, Omar and Rimlinger, Hugo and Dreyfus, Milo and Gouel, Matthieu and Vermeulen, Kevin},
    title = {{Replication: Towards a Publicly Available Internet Scale IP Geolocation Dataset}},
    year = {2023},
    publisher = {ACM},
    booktitle = {IMC},

}

@article{ipmap,
    author = {Du, Ben and Candela, Massimo and Huffaker, Bradley and Snoeren, Alex C. and claffy, kc},
    title = {{RIPE IPmap Active Geolocation: Mechanism and Performance Evaluation}},
    year = {2020},
    issue_date = {April 2020},
    publisher = {ACM},
    
    volume = {50},
    number = {2},
    journal = {SIGCOMM Comput. Commun. Rev.},
    month = {may},

    numpages = {8},
}

@INPROCEEDINGS{sav-sp,
  author={Lone, Qasim and Frik, Alisa and Luckie, Matthew and Korczyński, Maciej and van Eeten, Michel and Gañán, Carlos},
  booktitle={IEEE S\&P}, 
  title={{Deployment of Source Address Validation by Network Operators: A Randomized Control Trial}}, 
  year={2022},
  
}

@inproceedings{reverse-geolocation,
    author = {Gamero-Garrido, Alexander and Belding, Elizabeth and Choffnes, David},
    title = {{Using Reverse IP Geolocation to Identify Institutional Networks}},
    year = {2022},
    publisher = {ACM},
    booktitle = {IMC},

}

@inproceedings{ipvseeyou,
    author = {E. C. Rye and R. Beverly},
    booktitle = {IEEE S\&P},
    title = {{IPvSeeYou: Exploiting Leaked Identifiers in IPv6 for Street-Level Geolocation}},
    year = {2023},
    volume = {},
    publisher = {IEEE Computer Society},
}

@inproceedings{on-the-accuracy,
    author = {Livadariu, Ioana and Dreibholz, Thomas and Al-Selwi, Anas Saeed and Bryhni, Haakon and Lysne, Olav and Bj\o{}rnstad, Steinar and Elmokashfi, Ahmed},
    title = {{On the Accuracy of Country-Level IP Geolocation}},
    year = {2020},
    publisher = {ACM},
    booktitle = {ANRW},

}

@inproceedings{iot-location,
    author = {Bremler-Barr, Anat and Hay, David and Meyuhas, Bar and Danino, Shoham},
    title = {{It's Not Where You Are, It's Where You Are Registered: IoT Location Impact on MUD}},
    year = {2023},
    publisher = {ACM},
    booktitle = {ANRW},

}

@misc{iot-atlas,
    author = {Robert Kisteleki}, 
    title = {{RIPE Atlas and IoT}},
    howpublished = {\url{https://www.ripe.net/media/documents/20170921-iot-atlas.pdf}},
    year = 2017,
    month = sep,
}

@misc{data-retention,
    author = {Robert Kisteleki and Paul de Weerd}, 
    title = {{RIPE NCC Measurement Data Retention Principles}},
    howpublished = {\url{https://labs.ripe.net/author/kistel/ripe-ncc-measurement-data-retention-principles/}},
    year = 2023,
    month = nov,
}

@misc{ring,
  author = {{NLNOG}},
  title = {{Introduction - NLNOG RING}},
  howpublished = {\url{https://ring.nlnog.net}},
  year = 2024,
}

@misc{forbes-estimation,
  author = {Abram Brown},
  title = {{Facebook Lost About \$65 Million During Hours-Long Outage}},
  howpublished = {\url{https://www.forbes.com/sites/abrambrown/2021/10/05/facebook-outage-lost-revenue/?sh=565a1a88231a}},
  year = 2021,
}

@misc{reducing-footprint,
  author = {Felipe Victolla Silveira},
  title = {{Reducing the RIPE NCC's Data Centre Footprint}},
  howpublished = {\url{https://labs.ripe.net/author/felipe_victolla_silveira/reducing-the-ripe-nccs-data-centre-footprint/}},
  year = 2024,
}

@misc{2018-campaign,
  author = {Alun Davies},
  title = {{2018 Campaign to Sponsor 10 RIPE Atlas Anchors}},
  howpublished = {\url{https://labs.ripe.net/author/alun_davies/2018-campaign-to-sponsor-10-ripe-atlas-anchors/}},
  year = 2018,
}

@misc{atlas-400-anchors,
    author = {Alun Davies}, 
    title = {{RIPE Atlas Anchors 400+}},
    howpublished = {\url{https://labs.ripe.net/author/alun\_davies/ripe-atlas-anchors-400/}},
    year = 2019,
    month = jan,
}

@misc{user-defined,
    author = {{RIPE Atlas}}, 
    title = {{Starting your own Measurements (User-defined Measurements)}},
    howpublished = {\url{https://atlas.ripe.net/docs/getting-started/user-defined-measurements.html}},
    year = 2025,
}

@misc{scholar,
    author = {{Google Scholar}}, 
    title = {{"RIPE Atlas"}},
    howpublished = {\url{https://scholar.google.com/scholar?hl=en\&as\_sdt=0\%2C5\&q="RIPE+Atlas"\&btnG=}},
    year = 2024,
    month = may,
}

@misc{troubleshooting,
    author = {Philip Homburg}, 
    title = {{Troubleshooting RIPE Atlas Probes: USB Sticks}},
    howpublished = {\url{https://labs.ripe.net/author/philip\_homburg/troubleshooting-ripe-atlas-probes-usb-sticks/}},
    year = 2016,
    month = jul,
}

@misc{builtin,
  title = {{Built-in Measurements}},
  author = {{RIPE Atlas}},
  howpublished = {\url{https://atlas.ripe.net/docs/built-in-measurements/}},
  year = 2024
}

@misc{dnsmon-visualising,
    title = {{Visualising DNS Issues with DNSMON}},
    author = {Massimo Candela},
    howpublished = {\url{https://labs.ripe.net/author/massimo\_candela/visualising-dns-issues-with-dnsmon/}},
    year = 2024
}

@misc{domainmon,
    title = {{RIPE Atlas: DomainMON is Here}},
    author = {Suzanne Taylor},
    howpublished = {\url{https://labs.ripe.net/author/suzanne\_taylor\_muzzin/ripe-atlas-domainmon-is-here/}},
    year = 2015
}

@inproceedings{ooni,
  title = {{OONI: Open Observatory of Network Interference}},
  author = {{Arturo Filastò and Jacob Appelbaum}},
  booktitle = {2nd USENIX Workshop on Free and Open Communications on the Internet},
  pages = {1--8},
  publisher = {USENIX Association},
  year = {2012}
}

@techreport{proposal-240-1,
    number =    {draft-fuller-240space-02},
    type =      {Internet-Draft},
    institution =   {Internet Engineering Task Force},
    publisher = {Internet Engineering Task Force},
    note =      {Work in Progress},
    url =       {https://datatracker.ietf.org/doc/draft-fuller-240space/02/},
    author =    {Vince Fuller},
    title =     {{Reclassifying 240/4 as usable unicast address space}},
    pagetotal = 6,
    year =      2008,
    month =     mar,
    day =       25,
}

@techreport{proposal-240-2,
    number =    {draft-wilson-class-e-02},
    type =      {Internet-Draft},
    institution =   {Internet Engineering Task Force},
    publisher = {Internet Engineering Task Force},
    note =      {Work in Progress},
    url =       {https://datatracker.ietf.org/doc/draft-wilson-class-e/02/},
    author =    {Paul Wilson and George G. Michaelson and Geoff Huston},
    title =     {{Redesignation of 240/4 from "Future Use" to "Private Use"}},
    pagetotal = 7,
    year =      2008,
    month =     sep,
    day =       29,
}

@misc{240-4,
    author = {Qasim Lone and John Gilmore and Seth David Schoen and Dave Täht and Emile Aben}, 
    title = {{240/4 As Seen by RIPE Atlas}},
    howpublished = {\url{https://labs.ripe.net/author/qasim-lone/2404-as-seen-by-ripe-atlas/}},
    year = 2022,
    month = aug,
}

@misc{bundles,
    title = {{Measurement Bundles}},
    author = {{RIPE Atlas Community}},
    howpublished = {\url{https://github.com/RIPE-Atlas-Community/ripe-atlas-tips-and-tricks/wiki/Measurement-Bundles}},
    year = 2024
}

@misc{anti-patterns,
    title = {{Observations on RIPE Atlas API "Anti-Patterns"}},
    author = {Robert Kisteleki},
    howpublished = {\url{https://labs.ripe.net/author/kistel/observations-on-ripe-atlas-api-anti-patterns/}},
    year = 2023
}

@misc{ethics,
    title = {{Ethics of RIPE Atlas Measurements}},
    author = {Robert Kisteleki},
    howpublished = {\url{https://labs.ripe.net/author/kistel/ethics-of-ripe-atlas-measurements/}},
    year = 2016
}

@misc{terms,
    title = {{RIPE Atlas Service Terms and Conditions}},
    author = {{RIPE NCC}},
    howpublished = {\url{https://www.ripe.net/about-us/legal/ripe-atlas-service-terms-and-conditions/}},
    year = 2020
}

@misc{128,
    author = {Robert Kisteleki and Qasim Lone and Michel Stam}, 
    title = {{The Curious Case of Packets From ::}},
    howpublished = {\url{https://labs.ripe.net/author/kistel/the-curious-case-of-packets-from/}},
    year = 2022,
    month = may,
}

@misc{rfc4291,
    series =    {Request for Comments},
    number =    4291,
    howpublished =  {RFC 4291},
    publisher = {RFC Editor},
    url =       {https://www.rfc-editor.org/info/rfc4291},
    author =    {Dr. Steve E. Deering and Bob Hinden},
    title =     {{IP Version 6 Addressing Architecture}},
    pagetotal = 25,
    year =      2006,
    month =     feb,
}

@misc{pyasn,
  author = {Hadi Asghari},
  title = {{pyasn}},
  note = {\url{https://github.com/hadiasghari/pyasn}},
  year = {2020}
}

@misc{aggregators,
    title = {{The Role of Aggregators in RIPE Atlas}},
    author = {Robert Kisteleki},
    howpublished = {\url{https://labs.ripe.net/author/kistel/the-role-of-aggregators-in-ripe-atlas/}},
    year = 2023
}

@misc{fb-outage,
  author = {Santosh Janardhan},
  title = {{More details about the October 4 outage}},
  howpublished = {\url{https://engineering.fb.com/2021/10/05/networking-traffic/outage-details/}},
  year = 2021,
}

@misc{routeviews,
  author = {{University of Oregon}},
  title = {{Route Views Project}},
  note = {\url{http://www.routeviews.org/routeviews/}},
  year = {2020}
}

@inproceedings{snmp,
    author = {Albakour, Taha and Gasser, Oliver and Beverly, Robert and Smaragdakis, Georgios},
    title = {{Third Time's Not a Charm: Exploiting SNMPv3 for Router Fingerprinting}},
    year = {2021},
    publisher = {ACM},
    booktitle = {IMC},

}

@inproceedings{illuminating,
    author = {Albakour, Taha and Gasser, Oliver and Beverly, Robert and Smaragdakis, Georgios},
    title = {{Illuminating Router Vendor Diversity Within Providers and Along Network Paths}},
    year = {2023},
    publisher = {ACM},
    booktitle = {IMC},

}

@inproceedings{passive-analysis,
    author = {Enayet, Asma and Heidemann, John},
    title = {{Internet Outage Detection Using Passive Analysis}},
    year = {2022},
    publisher = {ACM},
    booktitle = {IMC},

}

@inproceedings{colocation-disruptions,
    author={Milolidakis, Alexandros and Fontugne, Romain and Dimitropoulos, Xenofontas},
    booktitle={IEEE INFOCOM}, 
    title={{Detecting Network Disruptions At Colocation Facilities}}, 
    year={2019},
    publisher = {{IEEE}},
    
}

@inproceedings{perceiving-anomalies,
    author={Jia, Yihao and Kuzmanovic, Aleksandar},
    booktitle={IEEE INFOCOM}, 
    title={{Perceiving Internet Anomalies via CDN Replica Shifts}}, 
    year={2019},
    
    publisher = {{IEEE}},
    
}

@inproceedings{disco,
    author = {Tomas Hlavacek and {\'{I}}talo Cunha and Yossi Gilad and Amir Herzberg and Ethan Katz{-}Bassett and Michael Schapira and Haya Schulmann},
    title = {{DISCO: Sidestepping RPKI's Deployment Barriers}},
    booktitle = {NDSS},
    publisher = {ISOC},
    year = {2020},
}

@inproceedings{rpki-validation,
    author = {Shulman, Haya and Vogel, Niklas and Waidner, Michael},
    title = {{Poster: Insights into Global Deployment of RPKI Validation}},
    year = {2022},
    publisher = {ACM},
    booktitle = {CCS},

}

@inproceedings{rovista,
    author = {Li, Weitong and Lin, Zhexiao and Ashiq, Md. Ishtiaq and Aben, Emile and Fontugne, Romain and Phokeer, Amreesh and Chung, Taejoong},
    title = {{RoVista: Measuring and Analyzing the Route Origin Validation (ROV) in RPKI}},
    year = {2023},
    publisher = {ACM},
    booktitle = {IMC},

}

@inproceedings{time-of-flight,
    
    author = {Fontugne, Romain and Phokeer, Amreesh and Pelsser, Cristel and Vermeulen, Kevin and Bush, Randy},
    booktitle = {PAM},
    publisher = {Springer Nature Switzerland},
    title = {{RPKI Time-of-Flight: Tracking Delays in the Management, Control, and Data Planes}},
    year = {2023}
}

@inproceedings{remote-peering,
    author={Mazzola, Fabricio and Marcos, Pedro and Castro, Ignacio and Luckie, Matthew and Barcellos, Marinho},
    title={{On the Latency Impact of Remote Peering}},
    booktitle={PAM},
    year={2022},
    publisher={Springer International Publishing},
    
    
}

@inproceedings{reassessing,
    author = {Lily Davisson and Joakim Jakovleski and Nhiem Ngo and Chau Pham and Joel Sommers},
    title = {{Reassessing the Constancy of End-to-End Internet Latency}},
    booktitle = {TMA},
    publisher = {{IFIP}},
    year = {2021},
}

@inproceedings{latency-characteristics,
    title={{Latency Characteristics of Edge and Cloud}},
    author={Charyyev, Batyr and Gunes, Mehmet Hadi},
    booktitle={TMA Lightning Talks Session},
    year={2020},
    publisher = {{IFIP}},
}

@inproceedings{unintended,
    author={Fanou, Rod{\'e}rick and Huffaker, Bradley and Mok, Ricky and Claffy, K. C.},
    title={{Unintended Consequences: Effects of Submarine Cable Deployment on Internet Routing}},
    booktitle={PAM},
    year={2020},
    publisher={Springer International Publishing},
    
    
}

@inproceedings{out-of-mind,
    author = {Liu, Shucheng and Bischof, Zachary S. and Madan, Ishaan and Chan, Peter K. and Bustamante, Fabi\'{a}n E.},
    title = {{Out of Sight, Not Out of Mind: A User-View on the Criticality of the Submarine Cable Network}},
    year = {2020},
    publisher = {ACM},
    booktitle = {IMC},

}

@inproceedings{starlink,
    author = {Michel, Fran\c{c}ois and Trevisan, Martino and Giordano, Danilo and Bonaventure, Olivier},
    title = {{A First Look at Starlink Performance}},
    year = {2022},
    publisher = {ACM},
    booktitle = {IMC},

}

@article{dissecting,
    author = {Raman, Aravindh and Varvello, Matteo and Chang, Hyunseok and Sastry, Nishanth and Zaki, Yasir},
    title = {{Dissecting the Performance of Satellite Network Operators}},
    year = {2023},
    issue_date = {December 2023},
    publisher = {ACM},
    
    volume = {1},
    number = {CoNEXT3},
    journal = {Proc. ACM Netw.},
    month = {nov},
    articleno = {15},
    numpages = {25},
}

@inproceedings{roll,
    author = {M\"{u}ller, Moritz and Thomas, Matthew and Wessels, Duane and Hardaker, Wes and Chung, Taejoong and Toorop, Willem and Rijswijk-Deij, Roland van},
    title = {{Roll, Roll, Roll your Root: A Comprehensive Analysis of the First Ever DNSSEC Root KSK Rollover}},
    year = {2019},
    publisher = {ACM},
    booktitle = {IMC},

}

@inproceedings{agility,
    author = {M\"{u}ller, Moritz and Toorop, Willem and Chung, Taejoong and Jansen, Jelte and van Rijswijk-Deij, Roland},
    title = {{The Reality of Algorithm Agility: Studying the DNSSEC Algorithm Life-Cycle}},
    year = {2020},
    publisher = {ACM},
    booktitle = {IMC},

}

@inproceedings{second-look,
    author = {Magnusson, Jonathan and M\"{u}ller, Moritz and Brunstrom, Anna and Pulls, Tobias},
    title = {{A Second Look at DNS QNAME Minimization}},
    year = {2023},
    publisher = {Springer-Verlag},
    booktitle = {PAM},

}

@inproceedings{first-look,
    author={de Vries, Wouter B. and Scheitle, Quirin and M{\"u}ller, Moritz and Toorop, Willem and Dolmans, Ralph and van Rijswijk-Deij, Roland},
    title={{A First Look at QNAME Minimization in the Domain Name System}},
    booktitle={PAM},
    year={2019},
    publisher={Springer International Publishing},
    
    
}

@inproceedings{cache-outside,
    author={Akhavan Niaki, Arian and Marczak, William and Farhoodi, Sahand and McGregor, Andrew and Gill, Phillipa and Weaver, Nicholas},
    title={{Cache Me Outside: A New Look at DNS Cache Probing}},
    booktitle={PAM},
    year={2021},
    publisher={Springer International Publishing},
    
    
}

@inproceedings{trufflehunter,
    author = {Randall, Audrey and Liu, Enze and Akiwate, Gautam and Padmanabhan, Ramakrishna and Voelker, Geoffrey M. and Savage, Stefan and Schulman, Aaron},
    title = {{Trufflehunter: Cache Snooping Rare Domains at Large Public DNS Resolvers}},
    year = {2020},
    publisher = {ACM},
    booktitle = {IMC},

}

@inproceedings{cache-me,
    author = {Moura, Giovane C. M. and Heidemann, John and Schmidt, Ricardo de O. and Hardaker, Wes},
    title = {{Cache Me If You Can: Effects of DNS Time-to-Live}},
    booktitle = {IMC},
    year = {2019},
    publisher = {ACM},
    

}

@inproceedings{dns-over-tcp,
    author = {Mao, Jiarun and Rabinovich, Michael and Schomp, Kyle},
    title = {{Assessing Support for DNS-over-TCP in the Wild}},
    year = {2022},
    publisher = {Springer-Verlag},
    booktitle = {PAM},
}

@inproceedings{dot-edge,
    author={Doan, Trinh Viet and Tsareva, Irina and Bajpai, Vaibhav},
    title={{Measuring DNS over TLS from the Edge: Adoption, Reliability, and Response Times}},
    booktitle={PAM},
    year={2021},
    publisher={Springer International Publishing},
    
    
}

@inproceedings{doh-world,
    author = {Chhabra, Rishabh and Murley, Paul and Kumar, Deepak and Bailey, Michael and Wang, Gang},
    title = {{Measuring DNS-over-HTTPS Performance Around the World}},
    year = {2021},
    publisher = {ACM},
    booktitle = {IMC},

}

@inproceedings{falling-bits,
    author={Moura, Giovane C. M. and M{\"u}ller, Moritz and Davids, Marco and Wullink, Maarten and Hesselman, Cristian},
    title={{Fragmentation, Truncation, and Timeouts: Are Large DNS Messages Falling to Bits?}},
    booktitle={PAM},
    year={2021},
    publisher={Springer International Publishing},
    
    
}

@inproceedings{apple-relay,
    author = {Sattler, Patrick and Aulbach, Juliane and Zirngibl, Johannes and Carle, Georg},
    title = {{Towards a Tectonic Traffic Shift? Investigating Apple's New Relay Network}},
    year = {2022},
    publisher = {ACM},
    booktitle = {IMC},

}

@inproceedings{tsuname,
    author = {Moura, Giovane C. M. and Castro, Sebastian and Heidemann, John and Hardaker, Wes},
    title = {{TsuNAME: Exploiting Misconfiguration and Vulnerability to DDoS DNS}},
    year = {2021},
    publisher = {ACM},
    booktitle = {IMC},

}

@inproceedings{ecs-behavior,
    author = {Al-Dalky, Rami and Rabinovich, Michael and Schomp, Kyle},
    title = {{A Look at the ECS Behavior of DNS Resolvers}},
    year = {2019},
    publisher = {ACM},
    booktitle = {IMC},

}

@inproceedings{edns-adoption,
    author = {Matt Calder and Xun Fan and Liang Zhu},
    title = {{A Cloud Provider's View of EDNS Client-Subnet Adoption}},
    booktitle = {TMA},
    publisher = {{IEEE}},
    year = {2019},
}

@inproceedings{peek,
    author = {Thiagarajan, Kedar and Kumar, Rashna and Bustamante, Fabi\'{a}n E.},
    title = {{Poster: A Peek Backstage: Organizations in DNS Resolver Hierarchies}},
    year = {2023},
    publisher = {ACM},
    booktitle = {SIGCOMM},

}

@inproceedings{managed-dns,
    author={Gao, Zhaoyu and Venkataramani, Arun},
    booktitle={IEEE INFOCOM}, 
    title={{Measuring Update Performance and Consistency Anomalies in Managed DNS Services}}, 
    year={2019},
    publisher = {{IEEE}},    
}

@inproceedings{curtain,
    author = {Jared M. Smith and Kyle Birkeland and Tyler McDaniel and Max Schuchard},
    title = {{Withdrawing the BGP Re-Routing Curtain: Understanding the Security Impact of BGP Poisoning through Real-World Measurements}},
    booktitle = {NDSS},
    publisher = {ISOC},
    year = {2020},
}

@inproceedings{zombies,
    author={Fontugne, Romain and Bautista, Esteban and Petrie, Colin and Nomura, Yutaro and Abry, Patrice and Goncalves, Paulo and Fukuda, Kensuke and Aben, Emile},
    title={{BGP Zombies: An Analysis of Beacons Stuck Routes}},
    booktitle={PAM},
    year={2019},
    publisher={Springer International Publishing},
    
    
}

@inproceedings{multipath,
    author={Li, Jie and Zhou, Shi and Giotsas, Vasileios},
    booktitle={IEEE INFOCOM Workshops}, 
    title={{Performance Analysis of Multipath BGP}}, 
    year={2021},
    publisher = {{IEEE}},
    
}

@inproceedings{flat-internet,
    author = {Arnold, Todd and He, Jia and Jiang, Weifan and Calder, Matt and Cunha, Italo and Giotsas, Vasileios and Katz-Bassett, Ethan},
    title = {{Cloud Provider Connectivity in the Flat Internet}},
    year = {2020},
    publisher = {ACM},
    booktitle = {IMC},

}

@inproceedings{igdb,
    author = {Anderson, Scott and Salamatian, Loqman and Bischof, Zachary S. and Dainotti, Alberto and Barford, Paul},
    title = {{iGDB: Connecting the Physical and Logical Layers of the Internet}},
    year = {2022},
    publisher = {ACM},
    booktitle = {IMC},

}

@inproceedings{reverse-traceroute,
    author = {Vermeulen, Kevin and Gurmericliler, Ege and Cunha, Italo and Choffnes, David and Katz-Bassett, Ethan},
    title = {{Internet Scale Reverse Traceroute}},
    year = {2022},
    publisher = {ACM},
    booktitle = {IMC},

}

@inproceedings{regional-topology,
    author = {Zhang, Zesen and Marder, Alexander and Mok, Ricky and Huffaker, Bradley and Luckie, Matthew and Claffy, K C and Schulman, Aaron},
    title = {{Inferring Regional Access Network Topologies: Methods and Applications}},
    year = {2021},
    publisher = {ACM},
    booktitle = {IMC},

}

@inproceedings{hoplets,
    author = {Raman, Prathy and Flores, Marcel},
    title = {{Building Out the Basics with Hoplets}},
    year = {2021},
    publisher = {Springer-Verlag},
    booktitle = {PAM},

}

@inproceedings{disagree,
    author = {Raffaele Sommese and Giovane C. M. Moura and Mattijs Jonker and Roland van Rijswijk{-}Deij and Alberto Dainotti and Kimberly C. Claffy and Anna Sperotto},
    title = {{When Parents and Children Disagree: Diving into DNS Delegation Inconsistency}},
    booktitle = {PAM},
    publisher = {Springer},
    year = {2020},
}

@inproceedings{metatrace,
    author = {Gouel, Matthieu and Darwich, Omar and Mouchet, Maxime and Vermeulen, Kevin},
    title = {{Poster: Towards a Publicly Available Framework to Process Traceroutes with MetaTrace}},
    year = {2023},
    publisher = {ACM},
    booktitle = {IMC},

}

@inproceedings{rrr,
    author = {Giotsas, Vasileios and Koch, Thomas and Fazzion, Elverton and Cunha, \'{I}talo and Calder, Matt and Madhyastha, Harsha V. and Katz-Bassett, Ethan},
    title = {{Reduce, Reuse, Recycle: Repurposing Existing Measurements to Identify Stale Traceroutes}},
    year = {2020},
    publisher = {ACM},
    booktitle = {IMC},

}

@inproceedings{loops,
    author={Alaraj, Abdulrahman and Bock, Kevin and Levin, Dave and Wustrow, Eric},
    title={{A Global Measurement of Routing Loops on the Internet}},
    booktitle={PAM},
    year={2023},
    publisher={Springer Nature Switzerland},
    
    
}

@inproceedings{cdn-routing,
    author = {Zhu, Jiangchen and Vermeulen, Kevin and Cunha, Italo and Katz-Bassett, Ethan and Calder, Matt},
    title = {{The Best of Both Worlds: High Availability CDN Routing Without Compromising Control}},
    year = {2022},
    publisher = {ACM},
    booktitle = {IMC},

}

@inproceedings{last-mile,
    author = {Fontugne, Romain and Shah, Anant and Cho, Kenjiro},
    title = {{Persistent Last-mile Congestion: Not so Uncommon}},
    year = {2020},
    publisher = {ACM},
    booktitle = {IMC},

}

@article{divided-edge,
    author = {Martin, Noah and Dogar, Fahad},
    title = {{Divided at the Edge - Measuring Performance and the Digital Divide of Cloud Edge Data Centers}},
    year = {2023},
    issue_date = {December 2023},
    publisher = {ACM},
    
    volume = {1},
    number = {CoNEXT3},
    journal = {Proc. ACM Netw.},
    month = {nov},
    articleno = {16},
    numpages = {23},
}

@inproceedings{painter,
    author = {Koch, Thomas and Yu, Shuyue and Agarwal, Sharad and Katz-Bassett, Ethan and Beckett, Ryan},
    title = {{PAINTER: Ingress Traffic Engineering and Routing for Enterprise Cloud Networks}},
    year = {2023},
    publisher = {ACM},
    booktitle = {SIGCOMM},

}

@inproceedings{youtube,
    author={Kiedanski, Diego and Nogueira, Mateo and Grampín, Eduardo},
    booktitle={IEEE INFOCOM Workshops}, 
    title={{Youtube Traffic From the Perspective of a Developing Country: the Case of Uruguay}}, 
    year={2019},
    publisher = {{IEEE}},   
}

@inproceedings{timeseries,
    author={Mouchet, Maxime and Vaton, Sandrine and Chonavel, Thierry},
    booktitle={IEEE INFOCOM Workshops}, 
    title={{Poster Abstract: A Flexible Infinite HMM Model for Accurate Characterization and Segmentation of RTT Timeseries}}, 
    publisher = {{IEEE}},
    year={2019},
}

@inproceedings{middle-east,
    author={Candela, Massimo and Gregori, Enrico and Luconi, Valerio and Vecchio, Alessio},
    booktitle={IEEE INFOCOM Workshops}, 
    title={{Dissecting the Speed-of-Internet of Middle East}}, 
    year={2019},
    publisher = {{IEEE}},
    
    
}

@inproceedings{private-wan,
    author={Arnold, Todd and Gürmeriçliler, Ege and Essig, Georgia and Gupta, Arpit and Calder, Matt and Giotsas, Vasileios and Katz-Bassett, Ethan},
    booktitle={IEEE INFOCOM}, 
    title={{(How Much) Does a Private WAN Improve Cloud Performance?}}, 
    year={2020},
    publisher = {{IEEE}},
    
}

@inproceedings{6hit,
    author={Hou, Bingnan and Cai, Zhiping and Wu, Kui and Su, Jinshu and Xiong, Yinqiao},
    booktitle={IEEE INFOCOM}, 
    title={{6Hit: A Reinforcement Learning-based Approach to Target Generation for Internet-wide IPv6 Scanning}}, 
    year={2021},
    publisher = {{IEEE}},
    
}

@inproceedings{home-hijacking,
    author = {Randall, Audrey and Liu, Enze and Padmanabhan, Ramakrishna and Akiwate, Gautam and Voelker, Geoffrey M. and Savage, Stefan and Schulman, Aaron},
    title = {{Home is Where the Hijacking Is: Understanding DNS Interception by Residential Routers}},
    year = {2021},
    publisher = {ACM},
    booktitle = {IMC},

}

@inproceedings{cryptocurrency,
    author = {Loe, Angelique Faye and Quaglia, Elizabeth Anne},
    title = {{You Shall Not Join: A Measurement Study of Cryptocurrency Peer-to-Peer Bootstrapping Techniques}},
    year = {2019},
    publisher = {ACM},
    booktitle = {CCS},

}

@inproceedings{selective-tampetring,
    author={Gamero-Garrido, Alexander and Carisimo, Esteban and Hao, Shuai and Huffaker, Bradley and Snoeren, Alex C. and Dainotti, Alberto},
    title={{Quantifying Nations' Exposure to Traffic Observation and Selective Tampering}},
    booktitle={PAM},
    year={2022},
    publisher={Springer International Publishing},
    
    
}

@inproceedings{kazakhstan,
    author = {Raman, Ram Sundara and Evdokimov, Leonid and Wurstrow, Eric and Halderman, J. Alex and Ensafi, Roya},
    title = {{Investigating Large Scale HTTPS Interception in Kazakhstan}},
    year = {2020},
    publisher = {ACM},
    booktitle = {IMC},
}

@inproceedings{netshuffle,
    author = {P. Kon and A. Gattani and D. Saharia and T. Cao and D. Barradas and A. Chen and M. Sherr and B. E. Ujcich},
    booktitle = {IEEE S\&P},
    title = {{NetShuffle: Circumventing Censorship with Shuffle Proxies at the Edge}},
    year = {2024},
    issn = {2375-1207},
    publisher = {IEEE Computer Society},
    month = {may}
}

@inproceedings{blocklisting,
    author = {Ramanathan, Sivaramakrishnan and Hossain, Anushah and Mirkovic, Jelena and Yu, Minlan and Afroz, Sadia},
    title = {{Quantifying the Impact of Blocklisting in the Age of Address Reuse}},
    year = {2020},
    publisher = {ACM},
    booktitle = {IMC},

}

@inproceedings{rov-mi,
    author = {Wenqi Chen and Zhiliang Wang and Dongqi Han and Chenxin Duan and Xia Yin and Jiahai Yang and Xingang Shi},
    title = {{ROV-MI: Large-Scale, Accurate and Efficient Measurement of ROV Deployment}},
    booktitle = {NDSS},
    publisher = {ISOC},
    year = {2022},
    
}

@inproceedings{inter-city,
    author = {Selim Ozcan and Ioana Livadariu and Georgios Smaragdakis and Carsten Griwodz},
    title = {{Longitudinal Analysis of Inter-City Network Delays}},
    booktitle = {TMA},
    publisher = {{IEEE}},
    year = {2023},
}

@inproceedings{revisiting-rpki,
    author = {Nils Rodday and {\'{I}}talo S. Cunha and Randy Bush and Ethan Katz{-}Bassett and Gabi Dreo Rodosek and Thomas C. Schmidt and Matthias W{\"{a}}hlisch},
    title = {{Revisiting RPKI Route Origin Validation on the Data Plane}},
    booktitle = {TMA},
    publisher = {{IFIP}},
    year = {2021},
}

@inproceedings{bidirectional-anycast,
    author = {Lan Wei and Marcel Flores and Harkeerat Bedi and John S. Heidemann},
    title = {{Bidirectional Anycast/Unicast Probing (BAUP): Optimizing CDN Anycast}},
    booktitle = {TMA},
    publisher = {{IFIP}},
    year = {2020},
}

@inproceedings{debogonising,
    author = {Stephen D. Strowes and Emile Aben and Ren{\'{e}} Wilhelm and Florian Obser and Riccardo Stagni and Agustin Formoso},
    title = {{Debogonising 2a10: : /12: Analysis of One Week's Visibility of a New /12}},
    booktitle = {TMA},
    publisher = {{IFIP}},
    year = {2020},
}

@inproceedings{negative-caching,
    author = {Shafir, Lior and Afek, Yehuda and Bremler-Barr, Anat and Peleg, Neta and Sabag, Matan},
    title = {{DNS Negative Caching in the Wild}},
    year = {2019},
    publisher = {ACM},
    booktitle = {SIGCOMM Posters and Demos},

}

@inproceedings{count-bots,
    author = {Leon B{\"{o}}ck and Dave Levin and Ramakrishna Padmanabhan and Christian Doerr and Max M{\"{u}}hlh{\"{a}}user},
    title = {{How to Count Bots in Longitudinal Datasets of IP Addresses}},
    booktitle = {NDSS},
    publisher = {ISOC},
    year = {2023},
}

@inproceedings{bias-in-platforms,
    author={Sermpezis, Pavlos and Prehn, Lars and Kostoglou, Sofia and Flores, Marcel and Vakali, Athena and Aben, Emile},
    booktitle={TMA}, 
    title={{Bias in Internet Measurement Platforms}}, 
    year={2023},
    publisher = {{IEEE}},
}

@inproceedings{metis,
    author = {Malte Appel and Emile Aben and Romain Fontugne},
    title = {{Metis: Better Atlas Vantage Point Selection for Everyone}},
    booktitle = {TMA},
    publisher = {{IFIP}},
    year = {2022},
}

@inproceedings{mirai,
    author = {Manos Antonakakis and Tim April and Michael Bailey and Matt Bernhard and Elie Bursztein and Jaime Cochran and Zakir Durumeric and J. Alex Halderman and Luca Invernizzi and Michalis Kallitsis and Deepak Kumar and Chaz Lever and Zane Ma and Joshua Mason and Damian Menscher and Chad Seaman and Nick Sullivan and Kurt Thomas and Yi Zhou},
    title = {{Understanding the Mirai Botnet}},
    booktitle = {USENIX Security},
    year = {2017},
    publisher = {USENIX Association},
    month = aug
}

@inproceedings{longitudinal,
    author = {Taejoong Chung and Roland van Rijswijk-Deij and Balakrishnan Chandrasekaran and David Choffnes and Dave Levin and Bruce M. Maggs and Alan Mislove and Christo Wilson},
    title = {{A Longitudinal, End-to-End View of the DNSSEC Ecosystem}},
    booktitle = {USENIX Security},
    year = {2017},
    publisher = {USENIX Association},
}

@inproceedings{tunneling,
    author = {Chung, Taejoong and Choffnes, David and Mislove, Alan},
    title = {{Tunneling for Transparency: A Large-Scale Analysis of End-to-End Violations in the Internet}},
    year = {2016},
    publisher = {ACM},
    booktitle = {IMC},
}

@inproceedings{resident-evil,
    author={Mi, Xianghang and Feng, Xuan and Liao, Xiaojing and Liu, Baojun and Wang, XiaoFeng and Qian, Feng and Li, Zhou and Alrwais, Sumayah and Sun, Limin and Liu, Ying},
    booktitle={IEEE S\&P}, 
    title={{Resident Evil: Understanding Residential IP Proxy as a Dark Service}}, 
    year={2019},
    
}

@inproceedings{how-great,
    author = {Nguyen Phong Hoang and Arian Akhavan Niaki and Jakub Dalek and Jeffrey Knockel and Pellaeon Lin and Bill Marczak and Masashi Crete-Nishihata and Phillipa Gill and Michalis Polychronakis},
    title = {{How Great is the Great Firewall? Measuring Chinas DNS Censorship}},
    booktitle = {USENIX Security},
    year = {2021},
    publisher = {USENIX Association},
    month = aug
}

@article{menlo,
    author = {M. Bailey and E. Kenneally and D. Maughan and D. Dittrich},
    journal = {IEEE Security \& Privacy},
    title = {{The Menlo Report}},
    year = {2012},
    volume = {10},
    number = {02},
    publisher = {IEEE Computer Society},
    month = mar
}

@inproceedings{ttl-violations,
    author = {Bhowmick, Protick and Ashiq, Md. Ishtiaq and Deccio, Casey and Chung, Taejoong},
    title = {{TTL Violation of  DNS Resolvers in the Wild}},
    year = {2023},
    publisher = {Springer-Verlag},
    booktitle = {PAM},

}

@article{squatters,
    author = {Salamatian, Loqman and Arnold, Todd and Cunha, \'{I}talo and Zhu, Jiangchen and Zhang, Yunfan and Katz-Bassett, Ethan and Calder, Matt},
    title = {{Who Squats IPv4 Addresses?}},
    year = {2023},
    issue_date = {January 2023},
    publisher = {ACM},
    
    volume = {53},
    number = {1},
    journal = {SIGCOMM Comput. Commun. Rev.},
    month = {apr},

}

@InProceedings{DeVries2015,
  author    = {De Vries, Wouter and Santanna, José Jair and Sperotto, Anna and Pras, Aiko},
  booktitle = {Intelligent Mechanisms for Network Configuration and Security},
  title     = {{How Asymmetric Is the Internet?: A Study to Support the Use of Traceroute}},
  year      = {2015},
  publisher = {Springer International Publishing},
}

@InProceedings{YihuaHe2005,
  author    = {Yihua He and Michalis Faloutsos and Srikanth Krishnamurthy and Bradley Huffaker},
  booktitle = {IEEE GLOBECOM},
  title     = {{On Routing Asymmetry in the Internet}},
  year      = {2005},
  publisher = {IEEE},
}

\end{document}